\documentclass[a4paper,12pt]{articlewp}
\usepackage{amsmath}%
\usepackage{amsfonts}%
\usepackage{amssymb}%
\usepackage{mathptmx}
\usepackage{rotating}
\usepackage{bigstrut}
\usepackage{multirow}
\usepackage{booktabs}
\usepackage{threeparttable}
\usepackage{setspace}%% set space between lines
\usepackage{type1cm} %% for scaling font size arbitarily        
\usepackage{makeidx}         % allows index generation
\usepackage{graphicx}        % standard LaTeX graphics tool
\usepackage{multicol}        % used for the two-column index
\usepackage[section]{placeins}

\usepackage{natbib}
\usepackage[hidelinks, hyperfootnotes=false]{hyperref}
\usepackage[flushmargin]{footmisc}% conflicts with hyperref

\usepackage{geometry}
\usepackage{changepage}

\usepackage{dcolumn}
\newcolumntype{.}{D{.}{.}{-1}}

\usepackage{caption}
\DeclareCaptionFont{white}{\color{white}}
\DeclareCaptionFormat{listing}{\colorbox{dkgray}{\parbox{\textwidth}{#1#2#3}}}
\usepackage{listings}
\usepackage{color}

\usepackage{courier} 
\definecolor{dkgreen}{rgb}{0,0.6,0}
\definecolor{gray}{rgb}{0.5,0.5,0.5}
\definecolor{lightgray}{rgb}{0.85,0.85,0.85}
\definecolor{dkgray}{rgb}{0.4,0.4,0.4}
\definecolor{mauve}{rgb}{0.58,0,0.82}
\usepackage{lmodern} % this one is good and should consider for thesis doc
\usepackage[T1]{fontenc}

\usepackage{amssymb,amsmath}
\usepackage{lineno}

\usepackage{authblk}

\title{Estimation of firm-level productivity changes in the Indian power sector:\\ Disentangling unobserved heterogeneity by a transformed fixed-effect stochastic frontier model\thanks{An earlier version of this paper was presented at the doctoral research workshop at Indian Institute of Management Bangalore and we are grateful to seminar particiapants and V Ranganathan for constructive comments. We are grateful to Hun-Jen Wang for providing software code and test data that helped us to develop the software procedure used in this paper. We thank K Kasturirangan and U. K. Sharma of the Planning Commission, Government of India for guidance on India's power sector policy. We thank U. R. Prasad and Rahul Banerjee of the Central Electricity Regulatory Commission and Kapil Dev of Power Exchange India Limited for comments on electricity markets and power sector reforms in India. Finally, we gratefully acknowledge financial support from SAP Labs India scholarship.}
}

\author[1]{\normalsize Anish Sugathan\thanks{Corresponding Author: \href{mailto:anish.s08@iimb.ernet.in}{\nolinkurl{anish.s08@iimb.ernet.in}}}}
\author[2]{\normalsize Deepak Malghan}
\author[1]{\normalsize S. Chandrashekar}
\author[1]{\normalsize Deepak K. Sinha}
\affil[1]{\small Corporate Strategy and Policy Area, Indian Institute of Management Bangalore}
\affil[2]{\small Public Policy Area, Indian Institute of Management Bangalore}

\begin{document}
\date{}% removes the defalut date in title
\maketitle

\begin{abstract}
%% Text of abstract
We measure firm-level productivity changes in the Indian electricity sector during a period that witnessed several pro-market regulatory changes. Using information collected from multiple sources we construct a unique panel of generating firms and transmission and distribution utilities spanning the years 2000 to 2009. We employ a recently developed improvement in the Stochastic Frontier panel method that allows controlling for time-invariant unobserved heterogeneity. Using the method we jointly estimate inefficiency and exogenous determinants of inefficiency. We estimate a flexible translog production model and compute decomposition of productivity into components of changes in technology, efficiency, scale and price effect. During this period, especially post Electricity Act 2003, we observed a general decline in firm-level productivity at the mean rate of $-1.6\%$ per year. A positive and large technical change is observed in the sector at the rate of $8\%$ per year, attributable possibly to newer capacity addition. Except for smaller gas based generators, inefficiency is observed to be increasing at the mean rate of $3.1\%$ per year in the sector. Consistent with extant findings we also document no significant impact of un-bundling on firm-level efficiency.



\end{abstract}

%\begin{keyword}
%%% keywords here, in the form: keyword \sep keyword
%
%%% MSC codes here, in the form: \MSC code \sep code
%%% or \MSC[2008] code \sep code (2000 is the default)
%
%Power Sector Reforms \sep Productivity Changes \sep Stochastic Frontier Analysis
%
%\end{keyword}

%\end{frontmatter}

%% Chapter 02
%% 
%\chapter{Estimation of firm-level productivity changes in the Indian power sector: Disentangling unobserved heterogeneity by a transformed fixed-effect stochastic frontier model}
\newpage
%\doublespacing
\onehalfspacing

\section{Introduction}

In the past two decades the Indian power sector has witnessed structural reforms through several landmark regulatory changes. Along the lines of power sector reforms elsewhere internationally, primarily reforms in India also emphasized un-bundling of vertically integrated utilities into functionally separate entities dealing with generation (production) and transmission and distribution (T\&D) (service). The reforms also attempted to attract private capital to the sector. The primary policy objectives of initiating reforms in the sector were anticipated efficiency gains and cost reduction. Therefore, an empirical assessment of firm-level productivity changes in the generators and T\&D utilities shall reveal the extent to which reforms have influenced the sector in the intended direction. However, pan-India measurement of productivity changes across the power sector value-chain poses two key challenges. First, firm-level heterogeneity due to diversity in geography, local regulations,technology employed and other unobserved factors makes pan-sector (and pan-India) measurements prone to omission-bias. Second, due to relatively lax regulatory requirements of central collection and maintenance of firm-level operating data in India, productivity measurements have to rely on data collated from multiple sources or estimated from aggregate numbers. Hence, a majority of extant research investigating efficiency changes in firms in the Indian power sector have have been confined in scope to specific geography, firm or technology.\\

In this context we attempt to estimate pan-India firm-level productivity changes in the generating firms, T\&D utilities and a few remaining vertically integrated utilities. Our empirical strategy to measure pan-sector efficiency change is to structurally control for firm-level heterogeneity.\footnote{After controlling for firm-level fixed effects, two important endogeneity issues still remain unresolved. First, the problem of simultaneity, that relates to the fact that managers may adjust the firms' output in accordance to the observed efficiency of inputs. Second, the problem of selection bias, that results from the fact that negative productivity shocks may drive firms to exit and due to absence of information on non-existent firms the observations only represents a truncated distribution. These issues are not unique to our research setting, as \cite{Fabrizio2007} points out, studies of the electricity industry have typically not treated both these issues. However in our setting, firms are predominantly government owned and are not forced to exit due to low profitability. In addition, electricity production in India has remained in short supply historically, therefore the opportunity to cut back on output due to inefficiency of inputs hardly exists. Hence, unobserved heterogeneity remains the larger econometric issue that we proceed to address in our empirical work.} Causal inference is crucial specifically in studies attempting to attribute observed efficiency differences to explanatory factors. For instance, investigating the influence of un-bundling on the performance of Indian power sector, \cite{Cropper2011} employ the \mbox{difference-in-difference} econometric technique. The method adjusts for omitted variable bias caused due to unobserved variables aggregated at the state and time period level. Therefore by design, the method provides the regression coefficients on explanatory variables close to causal interpretation. In our setting to enable causal inference in panel SFA models that jointly estimates inefficiency and the exogenous determinants of inefficiency, \citep[models following][]{Battese1995}, it is critical to control for unobserved heterogeneity. In a recent comparative investigation of methods, \cite{Kopsakangas2011} measures inefficiencies in the Finnish electricity distribution utilities using five different SFA models. The study reports that models accounting for unobserved heterogeneity produce lower inefficiency measures and considerably different inefficiency rank orders. Thus, ignoring unobserved heterogeneity could result in confounded regression coefficients with severely limited causal interpretation.\\

\cite{Greene2005,Greene2005b} suggests two new methods for controlling unobserved heterogeneity in panel SFA models: the ``\emph{true fixed-effects}'' model and the ``\emph{true random-effects}'' model. The problem of identification\footnote{A well known problem with conventional fixed effects SFA models with the assumption of time-invariant inefficiency is that its not possible to distinguish inefficiency from unobserved heterogeneity captured by the fixed effect term \citep{Schmidt1984}.} is addressed in these newer models by structurally constraining the positive inefficiency term to be time-varying and the unobserved heterogeneity to be time-invariant. However, \cite{Wang2010} points out that the `\emph{true fixed-effects}'' SFA model suffers from the problem of incidental parameters \citep{Neyman1948, lancaster2000} that might contaminate estimates of other model parameters when simultaneous estimation of fixed effects and the inefficiency variance parameter is attempted. \cite{Wang2010} suggests a solution to this problem by developing a model that enables elimination of unobserved fixed-effect variables (either by \emph{within} or \emph{difference} transformation) prior to estimation of inefficiency.\\ 

In our study of the Indian power sector during the reforms period, we therefore employ the \cite{Wang2010} transformed SFA model to disentangle unobserved firm-level heterogeneity from technical inefficiency. We empirically investigate the nature of productivity changes in 98 firms operating in the Indian power sector during the period of 2000-2009. Our sample represents 51 generators , 38 transmission \& distribution firms and 9 vertically integrated utilities with a total of 542 firm-year observations. The unbalanced panel of sample contains observations of firms that are under the ownership of central government, state government and private investors. Our sample is fairly representative and accounts for $45.7\%$ of total electricity generated and $59.4\%$ of total electricity consumed in India during the period of 1999-2009. Using a flexible translog production specification we decompose the measure of productivity change into components of changes in technology, efficiency, scale and price effects. Based on the firm-level sample we estimate that post Electricity Act 2003 there had been no improvements in firm level productivity. In addition, bulk of the productivity change is attributable to technology change (newer capacity addition), wereas efficiency is observed to be generally declining. Further, we also show that reforms had a varying degree of influence on different entities contingent on its role in the value-chain, technology employed and ownership.

\section{Deregulation and Measuring Firm Level Productivity Change}
\label{sec:implication}
The global wave of restructuring since the early 1990's, systematically brought about significant changes in industry structure, ownership pattern and mode of regulation in the electricity sector in several countries. A common feature in many of these reforms initiatives is the dismantling of vertically integrated utilities, thus separating generation of electricity (production) from T\&D (service), such that coordination of demand-supply, post restructuring, happens over a specialized market based mechanism. It is suggested that the introduction of competition and market-based transactions in the sector is made on ex-ante anticipation of improvement in technical efficiency, reduction in operating costs and hence positive welfare gains \citep{Joskow1983}. \cite{Wolfram2005} argues that restructuring would lead to efficiency gains because of: (a) the new incentives faced by the incumbents to improve efficiencies, (b) takeover of inefficient older plants and the arrival of new entrants with newer technologies, and (c) competition driving efficient allocation of factors of production. Thus, in the restructured industry setting, ex-post measurement of efficiency improvements for firms across the electricity value-chain provides a basis for critical empirical scrutiny of the reforms policy\footnote{\cite{Fabrizio2007}, employing plant level data on generators in U.S., use the difference-in-difference method to find that introduction of market-based industry structure has led to modest medium-run cost efficiency gains. Similarly \cite{Newbery1997}, find that restructuring the U.K power sector led to gains in cost efficiency and reduction in emissions. An exhaustive survey of empirical studies on electricity sector reforms in the developing countries by \cite{Jamasb2005} finds that institutional development in the country and governance of the sector are key to success or failure of the reforms initiative. Similarly, in a cross country study spanning 36 developing countries, \cite{Zhang2008} finds that introduction of competition has relatively more significant impact than privatization in stimulating efficiency improvements}.\\ 

In this context, several extant studies investigate the influence of restructuring on firm-level productivity changes in the electricity sector. Among alternative methods, the non-parametric Data Envelopment Analysis (DEA) is a popular technique employed for the measurement of efficiency in the power sector, e.g. \cite{vaninsky2006}\footnote{A survey of DEA applications in energy and environmental studies can be found in reviews by \cite{Jamasb2000} and \cite{Zhou2008}.}. However, with access to panel data, the SFA method presents a natural way to incorporate information obtained from multiple observations of the production unit spread over time. And among other things, it also allows for a richer production specification and formal statistical testing of hypotheses \citep{Hjalmarsson1996}. SFA is used by \cite{Knittel2002}, using a Cobb-Douglas specification, and by \cite{Hiebert2002}, using the more flexible translog specification for measuring efficiency of U.S. power generators. \cite{Knittel2002} partly controls for firm-level fixed effects by incorporating plant vintage information in the SFA model. The study checks for influence of alternative regulatory schemes on inefficiency and finds that performance incentive based regulation is associated with improvement in plant-level efficiency. The SFA model used by \cite{Hiebert2002}, follows \cite{Battese1995} to jointly measure inefficiency and the influence of firm-level factors associated with the observed heterogeneity in measured inefficiency. The study identifies capacity utilization and ownership as factors influencing efficiency, overall mixed results on efficiency gains are seen across the restructured U.S. states.\\ 

The literature examining productive efficiency of the Indian electricity sector during the reforms phase is also growing. Scholars have used broadly three class of methods in their empirical work: the non-parametric DEA, parametric SFA, and other econometric specifications with a dependent plant-level efficiency variable (like plant load factor, or thermal efficiency). The DEA technique had been used to measure relative efficiencies by: \cite{Shrivastava2012} for thermal plants during 2008-2009, \cite{Yadav2010, Yadav2011} for divisions of distribution utility in the north Indian state of Uttarakhand, \cite{Thakur2006} for state owned utilities during the period 2001-2002, \cite{Chitkara1999} for thermal plants operated by NTPC,\footnote{National Thermal Power Corporation (NTPC), with majority ownership of the central government of India, is the single largest producer of electricity in India} and \cite{Singh1991} for state owned coal fired plants during 1986-1987. The parametric SFA method had been used by: \cite{Shanmugam2005} for 56 coal based plants for the period 1994-2001 (using a panel data specification), and by \cite{Khanna1999b} for 66 thermal plants during 1987-1990. Econometric specifications with plant-level efficiency as dependent variable had been used by: \cite{Khanna1999} on 63 coal based plants during 1987-1990 to check for regulatory and technology factors as determinants of efficiency, \cite{Cropper2011} follows the difference-in-difference method on a sample of 82 thermal power plants during 1994-2008, and \cite{Sen2010} use dynamic panel-data estimator with a sample of 19 states for 1991-2007. The later two studies specifically investigate the causal influence of restructuring on efficiency changes at the aggregate state \citep{Sen2010}, and plant \citep{Cropper2011} level. The research on the impact of restructuring reveals mixed outcomes. \cite{Sen2010} points out that there are substantial state-level differences in improvements attributable to heterogeneity in historic as well as political context. \citep{Cropper2011} finds that while un-bundling has not resulted in improvements in thermal efficiency, there has been improvements in capacity utilization ($+4.6\%$) and reduction in forced outages ($-2.9\%$).\\ 

We note that the extant research has focused either at the aggregate state-level or to level of generating plants. However, economically important decisions of investment in capacity, technology and choice of factor allocations are made at the level of the firm that operates several productive assets under its ownership and control. Especially post un-bundling of generation from distribution and transmission, the role of the firm as the decision making entity is more salient. Hence, in our empirical work we focus on the firm as the unit of analysis to measure dynamic changes in efficiency at the firm-level. We also trace the extent to which the changes are driven by factors such as ownership differences, vintage of assets, un-bundling status of the state in which the firm is operating and competition.     

\section{Indian Power Sector Reforms}
\label{sec:reform}
The Indian Electricity Act 1910, promulgated primarily to ensure safety, was
the earliest legislative attempt to regulate the working of the Indian electricity 
industry. Post independence, the Indian Constitution accorded \emph{concurrent
status} to the electricity sector, thus placing it simultaneously under the
responsibility of the central and state governments. Subsequently the
Electricity (Supply) Act 1948 came into effect that paved the way for the
formation of vertically integrated government owned agencies: the \emph{State
Electricity Boards} (SEBs), entrusted with the responsibility of generation,
transmission and distribution of electricity in the respective states. However,
the power sector continued to be characterized by capacity underutilization,
inefficient operations and financially imprudent pricing policies. This
consequently lead to chronic shortages, poor quality of supply, frequent
breakdowns and bankruptcy of the SEBs (World bank reference here). The genesis
of reforms in the power sector can be broadly traced to these deteriorating
conditions. \cite{Arun1998} presents a detailed discussion on the nature and
genesis of reforms, beginning the amendments to the Electricity Act 1910 and
Supply Act 1948 in the year 1991, which primarily opened up the sector to
private investments. Subsequently in 1998 the electricity Regulatory Commissions
Act was enacted resulting in the setting up of the Central Electricity
Regulatory Commission (CERC) and other state level regulatory bodies. While
primarily CERC was concerned with the regulation of firms owned and operated by
the central government, it also regulated coordination activities for entities
spanning multiple states. However these early attempts hardly made any
substantial impact on the growth and recovery of the power sector. In a critical
examination of this early phase of reforms \cite{Dsa1999}, and \cite{Dubash2001}
highlight that the piecemeal approach to reforms failed to rein in the
progressively languishing state of the power industry.\\

In contrast to these earlier fragmented reform attempts, a paradigm shift in the
power sector reform process was brought about by the legislation of the
Electricity Act 2003 \citep{GOI2003} on 10$^{th}$ July 2003. The Electricity Act
2003 replaced and consolidated the existing legislations aiming for substantial
structural changes in the Indian electricity industry. The salient features of
the Act included de-licensing of thermal and captive generation, de-licensing of
distribution in rural areas, open access to transmission, phased open access to
distribution, multiple licenses in distribution and recognition of electricity
trading as a distinct activity enabling the creation of electricity markets. A
detailed exposition of the implications of Electricity Act 2003 for generation,
transmission, distribution and electricity trading can be found in
\cite{Bhattacharyya2005, Ranganathan2004, Singh2006} and \cite{Thakur2005},
while several limitations of the Act are discussed in \cite{Sankar2004}.\\

While the reforms started in the early 1990s, structurally significant changes
where set in motion only after the enactment of the Electricity Act 2003,
especially in terms of potential to influence the technological efficiency of
firms operating in the electricity sector. The Act specifically articulates
intention to instill competition in the industry and outlines the framework to
transit from vertically integrated monopoly structure to a multi-buyer and
multi-seller model. With the establishment of wholesale electricity
market\footnote{The first Indian power exchange, Indian Energy Exchange Limited
(IEXL), became operational in June 2008 and Subsequently Power Exchange India
Limited (PXIL) came into existence in October 2008. These markets are in a
nascent stage of development with low transactions volumes. At present the two
exchanges together trade close to only 2\% of the total power generated in the
country. The functioning of these wholesale power markets is explored in a few
recent studies \citep{Shukla2011, Siddiqui2012}.}, the institutional framework
for a competitive industry structure got further augmented. Similar to
observations by \cite{Ranganathan2004} and \cite{Singh2010} we also expect that
the series of structural changes in the electricity sector, especially during
the decade starting year 2000, demonstrates potential to incentivize firms to
improve technological as well as operational efficiency. In addition, given wide
variation in ownership structure, local regulation and historical context, we
anticipate substantial pan India heterogeneity in firm-level productivity
outcome, in response to these institutional incentives. It is in this
overarching insitutional context that we empirically investigate a substantial
duration of the reform period (2000-2009), attempting to measure the extent of
overall productivity improvement in the sector and identify exogenous observable
factors responsible for firm-level heterogeneity in outcomes.
\section{Data and Method}
\label{sec:datameth}
\subsection{Data}
\label{subsec:data}
We create a sample dataset of Indian power generators and T\&D utilities for the period of 2000-2009. The sample represents about 46\% of total generation and about 60\% of total electricity consumption in India during the period. The sample spans across 19 states and represents ownership under central Government, state Government and private investors. We collect from multiple sources data on total electricity generated/distributed and the factors of production, aggregated at the firm level. Variable definition, unit of measurement and respective sources of data is summarized in table \ref{tab:vardef} \& \ref{tab:vardef01}. Power generating firms are classified as ``coal-based'', ``gas-based'' or ``mixed'' depending on the type of fuel consumed most. Firms with generating assets using  coal, gas and other sources with no one dominant fuel type is classified under the ``mixed'' category. Similarly, firms engaged only in T\&D function are classified as ``distribution utilities'' and firms operating generators as well as engages in  T\&D are classified as ``vertically integrated''. The distribution of firms across the various categories is described in table \ref{tab:02sample}. Summary statistics for all the variables is shown in table \ref{tab:02sum2} and the distribution of key input-output variables across categories of firms is shown in table \ref{tab:02sum1}. The unit of fuel input is normalized to energy equivalent GWhr units. From table \ref{tab:02sum1}, the ratio of electricity generated to fuel input shows an aggregate input-output efficiency of about $28\%$ and $26\%$ for coal and gas based generators respectively. Tranmission loss estimated from the distribution utilities is about $28\%$. These estimates of aggregate efficiencies conforms well with other estimates based on plant level measurements like \cite{CEA2008}. 

\subsection{Transformed fixed-effect stochastic frontier model}
\label{subsec:model}
%% SFA Model Section
%% \label{subsec:model}
\noindent We start by representing a primal stochastic production 
frontier using a deterministic kernel $f(x_{nit},t;\beta_{n})$ producing 
a scalar output $y_{it}$ as
\begin{equation}
y_{it}=f(x_{nit},t;\beta_{n}).exp(\epsilon_{it}),
\label{eq:01}
\end{equation}
\noindent for the $i^{th}$ producer $i=1,...,I$ during time period $t=1,...,T$ using 
inputs $x_{n}, n=1,...,N$, where $\epsilon_{it}$ represents producer specific time-varying 
stochastic inefficiency term. We choose the flexible translog form, developed in 
\cite{Christensen1971, Christensen1973}, to express the time-varying
stochastic production function in equation~(\ref{eq:01}). The translog production function 
is a local second-order approximation to any arbitrary production function, 
and thus displays several desirable properties for empirical estimation. The translog 
specification places no prior functional constrains on returns to scale, elasticity of 
substitution between inputs and homotheticity. \cite{Christensen1973}, 
discuss the aforementioned and  other related properties of the translog 
production function in detail. Additionally \cite{Diewert1976} shows the 
translog form to be ``exact'' for the Divisia index \citep{Divisia1925}. 
We shall use the Divisia index subsequently to estimate efficiency change 
and productivity change over the period of study. For our sample of 
unbalanced panel data on $I$ producers over $T$ time periods we assume that
the production function can be expressed in the following translog form
\begin{equation}
\begin{split}
\text{ln }E_{it}&=\alpha_{i}+\beta_{K}\text{ ln }K_{it}+\beta_{L}\text{ ln }L_{it}
+ \beta_{F}\text{ ln }F_{it} + \beta_{t}t\\
& \quad + \frac{1}{2}\beta_{KK}\text{ ln }K_{it}^{2} + \frac{1}{2}\beta_{LL}\text{ ln }L_{it}^{2}
+ \frac{1}{2}\beta_{FF}\text{ ln }F_{it}^{2}\\
& \quad + \beta_{KL}\text{ ln }K_{it}L_{it} + \beta_{KF}\text{ ln }K_{it}F_{it}
+ \beta_{LF}\text{ ln }L_{it}F_{it}\\
& \quad +\frac{1}{2}\beta_{tt}t^{2}+\beta_{Kt}t\text{ ln }K_{it} +\beta_{Lt}t\text{ ln }L_{it} 
+\beta_{Ft}t\text{ ln }F_{it}+ \epsilon_{it},
\end{split}
\label{eq:trlog}
\end{equation}  
\noindent We define the inefficiency term $\epsilon_{it}=v_{it}-u_{it}$, were $v_{it}\sim N(0,\sigma_{v}^{2})$ is the noise component and $u_{it}$ 
is the nonnegative stochastic technical inefficiency component.
Similar to \cite{kumbhakar2003}, in this translog specification as well, 
time ($t$) proxies technical change in the stochastic production frontier 
as well as represents technical efficiency change in the error component. 
Subsequently we shall impose distributional and model specification 
restrictions to econometrically disentangle the two effects. We attempt
to separate the firm specific unobserved 
heterogeneity, like \cite{Greene2005b}, by introducing the $\alpha_{i}$ 'fixed-effect' term
\footnote{Other formulations that treat both the firm-specific heterogeneity
     $\alpha_{i}$ as well as the the technical inefficiency error component
     $u_{i}$ to be time-invariant encounter a fundamental problem of identification.
     In such specifications (e.g.\cite{Pitt1981} and \cite{Schmidt1984}) the 
     time invariant term remain inseparable in the form $\alpha_{i}-u_{i}$. 
     However, with a time-varying inefficiency specification $u_{it}$,
     the presence of within group variation in the sample enables separate 
     analysis of unobserved heterogeneity and inefficiency. This separability
     between the two effects is limited by the extend to which technical inefficiency
     is time persistent. \cite{Greene2005b} analyses these issues in greater detail.}.
The consequent technical challenges in econometric estimation of such a specification
arise broadly due to, (a) the increased computational burden on account of introduction
of additional unknown parameters for estimation (one additional parameter for each firm in the sample in case of fixed-effect model)., (b)the problem of incidental parameters \citep{Neyman1948, lancaster2000} contaminates estimates of other model parameters when simultaneous estimation
of $\alpha_{i}$ and the inefficiency parameter is attempted (e.g. the ``true fixed-effect'' proposed in \cite{Greene2005b} and \cite{Greene2005}). The former of the two issues is addressed to some extent by recent developments in computer algorithms (one such algorithm is detailed 
in \cite{Greene2005}). However, in a sample with fixed $T$ and where $I\rightarrow \infty$, the later problem of incidental parameters results
in inconsistent estimates of the variance-covariance matrix \citep{Wang2010}.
Since, the inefficiency parameters of interest are contained in 
the variance-covariance matrix, we cannot afford to ignore inconsistency of 
estimates produced by a maximum likelihood estimator (MLE). \cite{Wang2010} 
propose a transformation for the panel stochastic frontier model that allows tractable
MLE estimation of the 'true fixed-effects' model. We follow this approach to
estimate the parameters of the stochastic production function we specify in 
equation(\ref{eq:trlog}).  Here MLE tractability is achieved by the use of
`scaling-property' \citep{Alvarez2006,Wang2002} to represent the inefficiency
component $u_{it}$ as a product of a positive non-stochastic time-varying function
and a stochastic time-invariant inefficiency term as  
\begin{align}
u_{it}&=h_{it}.u_{i}^{*},\\
h_{it}&=g(z_{kit}\delta_{k})
\label{eq:02}
\end{align}
\noindent where $u_{i}^{*} \sim N^{+}(0,\sigma_{u}^{2})$ is 
assumed to be non-negative half-normal and $z_{kit}$ represents $k=1,...,K$ 
exogenous and non-stochastic determinants of inefficiency. In this 
``composed error'' representation ($\epsilon_{it}=v_{it}-u_{it}$), 
the noise component $v_{it}$ is assumed to be iid and distributed 
independently of $u_{it}$. Further both $u_{i}^{*}$ and $v_{it}$ 
are assumed to be independent of $\{x_{nit},z_{kit}\}$ for all $T$ 
observations of the $i^{th}$ firm. The scaling property lends several
theoretically appealing properties enabling versatile model
specifications for empirical work, some of them are discussed in detail in 
\citep{Alvarez2006,Wang2002,Wang2010}. Specifically, we adopt  the 
'within' transformation of stochastic frontier models made tractable
by the scaling property. The 'within' transformation removes the 
individual fixed-effect (incidental parameter) $\alpha_{i}$ from
the model, thus inefficiency parameters can be estimated without
contamination due to the incidental parameter problem.\footnote{\cite{Wang2010} develop the 'first-difference' and 'within' transformation as two alternative approaches to eliminate the incidental variable. They also demonstrate that the log likelihood functions for both are equivalent. However, since the 'first-difference' transformation uses only $(T-1)$ observations from each panel in the sample we instead prefer to adopt the 
'within' transformation method for our empirical investigation.}
The 'within' transformation model that we use for MLE estimation is 
described in greater detail in the technical appendix (\ref{app:A}) to this paper. Furthermore
we specify the time-varying component of the inefficiency term, $h_{it}$, as
\begin{equation}
h_{it}=\text{exp}(z_{kit}\delta_{k}),
\label{eq:03}
\end{equation}
\noindent where, we investigate the influence of three different class of 
exogenous factors: (a) Vintage, proxied by the year of incorporation of the 
firm. (b) Ownership dummies, identifying a firm to central government,
or state government ownership class and private ownership is the reference class. (c)
External environmental factors, primarily the time since electricity sector
is unbundled in the state in which the firm has majority of operations and extent of competitiveness enabled by institutional conditions. (d) Time trend, all other dis-embodied 
factors proxied by a quadratic time specification. Then in equation(\ref{eq:03}), the inefficiency
determinant is specified as  
\begin{equation}
\begin{split}
 z_{kit}\delta_{k} &= time.\delta_{t} + time^{2}.\delta_{tt} 
								+ Vintage.\delta_{V} + Central~Gov.~Dummy.\delta_{CG}\\
		 &\quad			+ Private~Gov.~Dummy.\delta_{SG}\\						
		 &\quad			+ Unbundled~Dummy.\delta_{Udl}\\
		 &\quad			+ Competition.\delta_{Cmp},		 
\end{split}								
\label{eq:04}
\end{equation}

\subsection{Estimating productivity changes}
\label{subsec:prodch}  
%% Productivity Change Section
%% \label{subsec:prodch}  
Differentiating the production specification with respect to time, following \cite{kumbhakar2003}, yields various components of TFP change. The rate of shift in production frontier or technical change is given by
\begin{equation}
\Delta T = \frac{\partial \text{ ln } f(x,t;\beta)}{\partial t},
\end{equation} 
and the change in technical efficiency is obtained by
\begin{equation}
\Delta TE = -\frac{\partial u}{\partial t}.
\end{equation} 
The Divisia index of productivity change \citep{Divisia1925} is defined for a scalar output
as
\begin{equation}
\begin{split}
\dot{TFP}& =\frac{d\text{ ln } y}{d t} - \frac{d\text{ ln }X}{dt}\\
         & = \dot{y}-\dot{X} = \dot{y}-\sum_{n}(S_{n}\dot{x_{n}})
\end{split}
\label{eq:04}
\end{equation} 
Where $S_{n} = w_{n}x_{n}/\sum_{n}w_{n}x_{n}$ is the observed expenditure share
of the input $x_{n}$. Substituting for $\dot{y}$ in equation(\ref{eq:04}) obtained 
by totally differentiating equation(\ref{eq:01}) yields with minor algebraic manipulation the following decomposition of productivity change, 
\begin{equation}
\dot{TFP}=\Delta T + \Delta TE + (\Gamma-1)\sum_{n}(\frac{\gamma_{n}}{\Gamma})\dot{x_{n}}
+ \sum_{n}(\frac{\gamma_{n}}{\Gamma}-S_{n})\dot{x_{n}},     
\label{eq:05}
\end{equation}
where the elasticity of output with respect of input $x_{n}$ is defined as 
$\gamma_{n} = x_{n}\frac{\partial f}{\partial x_{n}}$. The returns to scale characterizing the 
production function is then expressed as $\Gamma=\sum_{n}(\gamma_{n})$. The third term in
equation (\ref{eq:05}),
\begin{equation}
\Psi = (\Gamma-1)\sum_{n}(\frac{\gamma_{n}}{\Gamma})\dot{x_{n}},
\end{equation}
represents the contribution of scale effects due to expansion or contraction 
of inputs towards total productivity change. Evidently under constant 
returns to scale ($\Gamma=1$) there is no contribution of scale effects.
However, under increasing/decreasing returns to scale ($\Gamma>1/\Gamma<1$)  
input expansion positively/negatively contributes to productivity change. The
allocative efficiency (or price effect) given by 
\begin{equation}
\Omega=\sum_{n}(\frac{\gamma_{n}}{\Gamma}-S_{n})\dot{x_{n}},
\end{equation}
\noindent represents productivity changes that are resulting from
factor prices being at deviance from their respective marginal
contribution to production. Thus, in case of factor prices reflecting 
perfect marginal costs ($\frac{\gamma_{n}}{\Gamma}-S_{n}=0$) 
the contribution due to price effect would be nil ($\Omega=0$). TFP change and its decomposition , derived in equation (\ref{eq:05}), can be computed using the parameter estimates of the production function in equation (\ref{eq:trlog}) as follows,
\begin{eqnarray}
\hat{\Delta T_{it}} & = &\hat{\beta}_{t}+ \hat{\beta}_{tt}t, \label{eqn:deltaT}\\
\hat{\Delta TE}_{it} & = & -\hat{u_{it}}\frac{d~h_{it}}{d t}   \approx-\hat{u_{it}}\frac{(h_{it}-h_{it_{-1}})}{(t-t_{-1})},\\
\hat{\gamma}_{nit} & = &\hat{\beta}_{n}+\sum_{k}\hat{\beta}_{nk}\text{ ln }\mathbf{x}_{it}+\hat{\beta}_{nt}t,~~n=1,\ldots,N,\\
\hat{\Gamma}_{it} &=& \sum_{n}\hat{\gamma}_{nit}
\end{eqnarray}

%\text{ln }E_{it}&=\alpha_{i}+\beta_{K}\text{ ln }K_{it}+\beta_{L}\text{ ln }L_{it}
%+ \beta_{F}\text{ ln }F_{it} + \beta_{t}t\\
%& \quad + \frac{1}{2}\beta_{KK}\text{ ln }K_{it}^{2} + \frac{1}{2}\beta_{LL}\text{ ln }L_{it}^{2}
%+ \frac{1}{2}\beta_{FF}\text{ ln }F_{it}^{2}\\
%& \quad + \beta_{KL}\text{ ln }K_{it}L_{it} + \beta_{KF}\text{ ln }K_{it}F_{it}
%+ \beta_{LF}\text{ ln }L_{it}F_{it}\\
%& \quad +\frac{1}{2}\beta_{tt}t^{2}+\beta_{Kt}t\text{ ln }K_{it} +\beta_{Lt}t\text{ ln }L_{it} 
%+\beta_{Ft}t\text{ ln }F_{it}+ \epsilon_{it},

\section{Results}
\label{sec:result}
Our empirical investigation is guided by two key objectives. First, to know the 
distribution and nature of productivity change in the power sector. Second, to identify 
the sources of inefficiency. We attempt to fulfill these objectives by first, jointly estimating inefficiency and the exogenous determinants of inefficiency. Then we decomposing the estimated TFP changes into constituents of: technical change, efficiency change, scale effects and price effects, to understand the nature of change.\\

For different classes of firms in the power sector, we estimate the primal production model, equation (\ref{eq:trlog}), using the transformed fixed-effects SFA method. We employ the Maximum Likelihood Estimator (MLE) technique to fit the model with empirical data. The estimated parameters of the model are shown in table (\ref{tab:model1}) and TFP and its decompositions are shown in table (\ref{tab:prod}). For all the estimated models, except `Mixed Generators', the inefficiency component, $ln(\sigma_{u})$, is significant and larger that the stochastic noise component $ln(\sigma_{v})$. Therefore, for these models the data shows existence of stochastic inefficiency different from noise. We observe that $\sum_{n} \beta_{n} \neq 1$ and $\beta_{nk} \neq \beta_{nt} \neq 0, ~\forall n,k$. Therefore, the production technology does not conforms to the linearly homogeneous and simpler Cobb-Douglas specification. This justifies our choice of the flexible translog specification and also implies that the scale component, $\gamma_{n}$, varies across firms and through time. 
We expect that the during this period TFP changed differently for the generators, T\&D and integrated utilities.

\subsection{Generators: Coal \& Gas}
For the coal based generators only ownership is significantly causing efficiency differences. We estimate the central Government owned generators to be about 57\% ($=1-e^{-0.835}$) less inefficienct than state Government or private owned ones. As expected the assets vintage influences reduction in inefficiency for newer plants, about 0.25\% ($=1-e^{-0.239/(10*(2008-1913))}$)reduction for assets newer by one year, however the effect is not statistically significant.No influence of compeition, un-bundling or time trend in inefficiency is observed. During this period the average per year TFP change observed is 11\%. We observe that in the period post Electricity Act 2003, after 2004, there had been increase in technical change (shift in frontier),13\% per year, while efficiency had been declining at $-7.5\%$ per year. The mean returns to scale, $\bar{\Gamma}=1.15$, indicates that coal based power generation shows increasing returns to scale.\\

\begin{figure}[h]
\centering
	\caption{Power Sector Technical Efficiency Time-Trend}
	\label{fig:EffTimeTrend}
	\centering
		\includegraphics[width=1.00\textwidth]{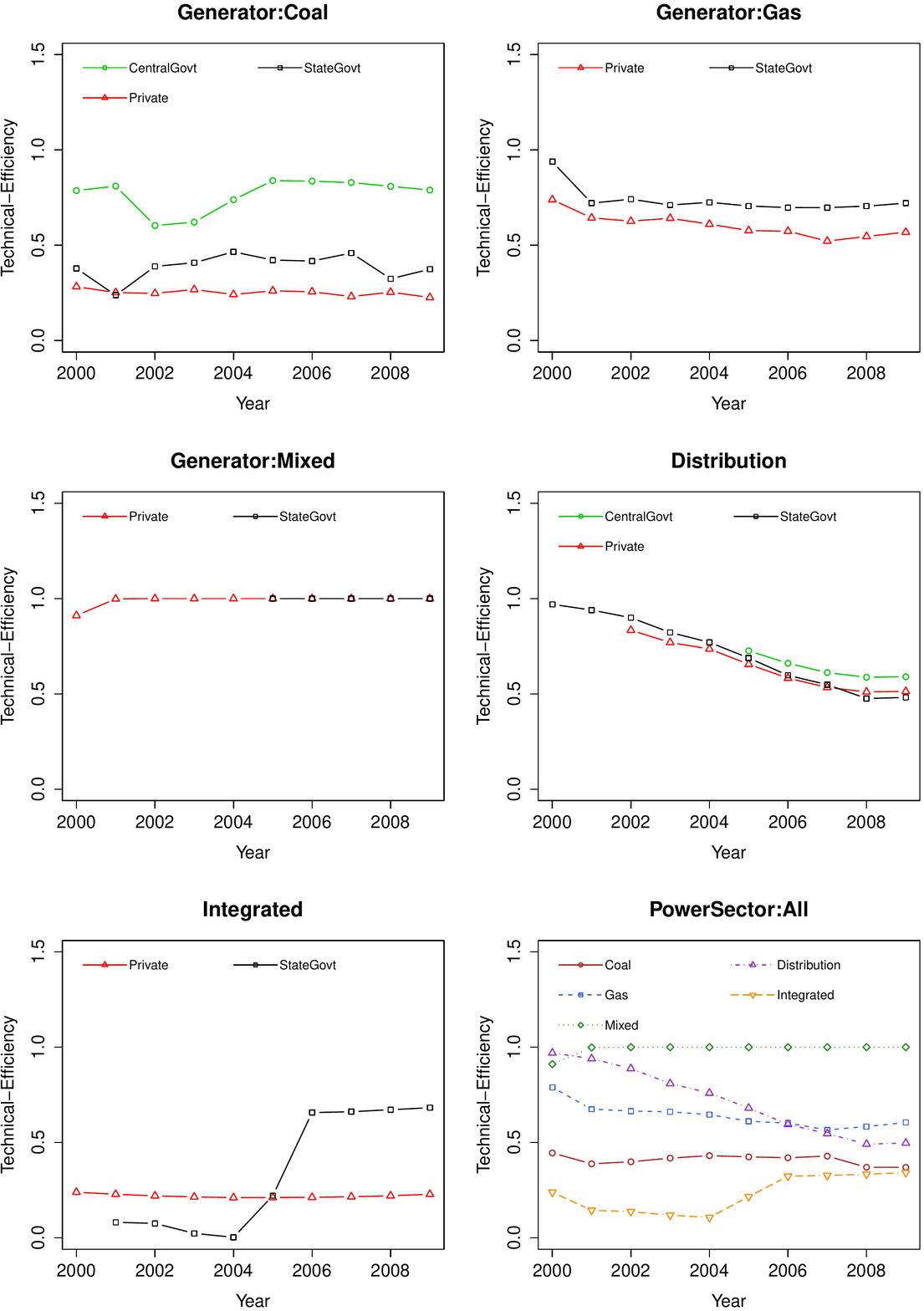}	
\end{figure}

For gas based generators, increased state-level competition is reducing inefficiency. Such that for every one index point increase in competition there is about 25\% reduction in inefficiency. Inefficiency also show a significant time-trend. The quadratic terms inidicate that inefficiency is increasing till the year 2007 and there is improvement subsequently (see figure (\ref{fig:EffTimeTrend})). No influence of asset vintage, un-bundling or ownership differences on inefficiency is observed. There is an average reduction in TFP of $-1.4\%$ per year, and the decline is mostly post year 2004. Post 2004, we observe that technical change has reduced from 12.8\% per year to 1.3\%, whereas efficiency change has improved from -7.6\% to -0.1\% per year. The mean returns to scale, $\bar{\Gamma}=0.56$, indicates that gas based power generation shows decreasing returns to scale. 

\subsection{T\&D \& Integrated Utilities}
For the T\&D utilities, assets vintage has a significant influence as seen by reduction in inefficiency for newer plants. About 1.6\% ($=1-e^{-1.496/(10*(2008-1913))}$) inefficiency reduction for assets newer by one year is observed. Inefficiency also show a significant time-trend. The quadratic terms inidicate that inefficiency is increasing at a reducing rate till the year 2008 and there is no decline subsequently (see figure (\ref{fig:EffTimeTrend})). No influence of competition, un-bundling or ownership differences on inefficiency is observed. TFP changed at a mean rate of 46\% per year. Post 2004, technical change reducted from 13.8\% to 8.4\% per year and efficiency change marginaly worsened from $-7.3\%$ to $-8.1\%$ per year.
The mean returns to scale, $\bar{\Gamma}=20$, indicates that T\&D firms show increasing returns to scale.\\

Only ownership is observed to be significantly associated with inefficiency differences for the integrated utilities. The state Government owned utilities are observed to be significantly inefficient compared to the private utilities. No influence of competition, un-bundling or time-trend in inefficiency is observed. TFP is changing at a mean rate of about $-11\%$ per year. Post 2004, technical change declined from 17.2\% to $-5.6\%$ per year, while efficiency improved from $-13.2\%$ to 3.3\% per year. The mean returns to scale, $\bar{\Gamma}=8$, indicates that integrated utilities show increasing returns to scale.

\section{Conclusion}
\label{sec:result}
Our results suggests that firm-level productivity in the Indian power sector has generally declined during the period of 2000-2009. We document that state-level un-bundling of the sector is not significantly associated with firm-level efficiency change.  Further, efficiency improvements attributable to increased competition is observed only in the case of smaller gas based generators. During the period post Electricity Act 2003, we observe positive technology change while simultaneously a decline in efficiency is observed for the coal based generators. Improvement in efficiency over time is observed only for gas based generators and integrated utilities, whereas T\&D firms show a decline in both technical change and efficiency.\\ 

Our results are consistent with earlier findings. For instance \cite{Cropper2011} finds no statistically significant improvement in thermal efficiencies post un-bundling while \cite{Sen2010, Cropper2011} find a significant improvement in plant-load factors (capacity utilization). We also observed a similar effect reflected in the increase in mean scale change effect from 1.8\% to 12\% per year post year 2004. However, contribution to TFP improvement from increased capcity utilization is offset partially by the declining efficiencies.\\ 

These results are indicative of the piecemeal approach to power sector reforms in India. The emphasis of reforms in India had been towards un-bundling of utilities and opening up the sector to private independent power producers. However, market for power remains under-developed, tariff reforms are not initiated and fuel remains short in supply. These anomalies are likely to creates skewed incentives for firms. The generators are governed by rate of return regulation and generally do not face retail competition. Therefore de-licensing investment in generation creates incentives for private investors to invest in large capital intensive projects. We observe this effect in the form of increased technical change only in the coal based generators, that is a result of increased investments in newer and larger plants. From a policy perspective, our results point towards the need for tariff reforms to encourage increased participation of independent power producers. For the T\&D firms controlled and low retail prices hardly makes-up for cost recovery and creates disincentive for private investments. For the larger generators there is lack of market incentives to reduce costs or improve efficiency, therefore strengthening the electricity markets and introduction of retail competition are possible policy alternatives to pursue. 

%\newpage
\singlespacing
\appendix
\section{Within Transformed SFA Model}
\label{app:A}
\nopagebreak[4]
%% Appendix A

\subsection{SFA Model Specification}
\noindent The within transformed SFA model \citep{Wang2010} used in this paper is described here. Consider an SFA model with the following general specifications$:$
\begin{eqnarray}
y_{it} & = &\alpha_{i}+\mathbf{x}_{nit}\boldsymbol\beta+\mathbf{\epsilon}_{it},
\hspace{1em} i=1,\ldots,I,\hspace{1em}t=1,\ldots,T,\hspace{1em}n=1,\ldots,N
\label{eqn:ineff1}\\
\epsilon_{it} & = &v_{it} + u_{it},\\
v_{it} &\sim& N(0,\sigma_{v}^{2}),\\
u_{it} &=& h_{it}~.~\bar{u}_{i},\\
h_{it} &=& f(\mathbf{z}_{kit}\mathbf{\delta}),\hspace{1em}k=1,\ldots,K\\
\bar{u}_{i} &\sim& N^{+}(\mu,\sigma_{u}^{2}).\label{eqn:ineff6}
\end{eqnarray}
Here, $\mathbf{x}_{nit}$ is a vector of $N$ production factor variables (or explanatory variables in general) and $\alpha_{it}$ represents unobserved fixed effect corresponding to the $i^{th}$ firm. $v_{it}\sim N(0,\sigma_{v}^{2})$ is the noise component and $\bar{u}_{it}$ 
is the nonnegative stochastic technical inefficiency component. While $\bar{u}_{it}$ is defined as a truncated normal distribution (Eq.\ref{eqn:ineff6}), in our model we set $\mu=0$ and assume a half-normal distribution for the inefficiency component. The vector $\mathbf{z}_{kit}$ represents $K$ exogenous variables determining inefficiency. 

\subsection{Transformed Specification}
The \emph{within transformation} is obtained by subtracting the sample mean of each panel
from every individual observation in the panel. The transformation, by de-meaning, removes
time-invariant fixed effects from the model. The model specification (Eq.\ref{eqn:ineff1}-\ref{eqn:ineff6}) post transformation may be represented as$:$
\begin{eqnarray}
\tilde{y}_{i*} & = &\alpha_{i}+\mathbf{\tilde{x}}_{ni*}\boldsymbol\beta+\mathbf{\tilde{\epsilon}}_{i*},
\label{eqn:tineff1}\\
\tilde{\epsilon}_{i*} & = &\tilde{v}_{i*} + \tilde{u}_{i*},\\
\tilde{v}_{i*} &\sim& \mathbf{N}(0,\Pi),\label{eqn:tineff3}\\
\tilde{u}_{i*} &=& \tilde{h}_{i*}~.~\bar{u}_{i},\\
\bar{u}_{i} &\sim& N^{+}(\mu,\sigma_{u}^{2}).\label{eqn:tineff5}
\end{eqnarray}
Here, we denote mean of individuals over the panel by $y_{i*} = (1/T)\Sigma^{T}_{t=1}y_{it}$, and the mean differenced value by $y_{it*}=y_{it}-y_{i*}$. The full panel as a vector stack is represented as $\tilde{y}_{i*} = (y_{i1},y_{i2},\ldots,y_{iT})'$. The variance-covariance matrix of $\tilde{v}_{i*}$ (Eqn. \ref{eqn:tineff3}) is
\begin{equation}
\label{eqn:mat}
\Pi = \begin{bmatrix}
\sigma^{2}_{v}(1-1/T) & \sigma^{2}_{v}(-1/T) &\cdots& \sigma^{2}_{v}(-1/T) \\
\sigma^{2}_{v}(-1/T) & \sigma^{2}_{v}(1-1/T) &\cdots& \sigma^{2}_{v}(-1/T) \\
\vdots & \vdots & \ddots & \vdots \\
\sigma^{2}_{v}(-1/T) & \sigma^{2}_{v}(-1/T) &\cdots& \sigma^{2}_{v}(1-1/T) \\
 \end{bmatrix} 
\end{equation}

\subsection{Log-Likelihood Function}
For the model described above, \cite{Wang2010} derives the marginal log-likelihood function of the $i^{th}$ panel as follows, 
\begin{equation}
\label{eqn:loglik}
\begin{split}
ln~L_{i} &= -\frac{1}{2}(T-1)ln(2\pi)-\frac{1}{2}(T-1)ln(\sigma^{2}_{v})-\frac{1}{2}\tilde{\epsilon}_{i*}'\Pi^{-}\tilde{\epsilon}_{i*}+\frac{1}{2}\left(\frac{\mu^{2}_{1}}{\sigma^{2}_{1}}-\frac{\mu^{2}}{\sigma^{2}}\right)\\
& +ln\left(\sigma_{1}\Phi\left(\frac{\mu_{1}}{\sigma_{1}}\right)\right) -ln\left(\sigma_{u}\Phi\left(\frac{\mu}{\sigma_{u}}\right)\right),
\end{split}
\end{equation}
where $\Pi^{-}$ is the generalized inverse of $\Pi$, $\phi$ the normal density function, $\Phi$ the cumulative density function and, 
\begin{equation}
\mu_{1}=\frac{\mu/\sigma_{u}^{2}-\tilde{\epsilon}_{i*}'\Pi^{-}\tilde{h}_{i*}}{\tilde{h}_{i*}'\Pi^{-}\tilde{h}_{i*} + 1/\sigma^{2}_{u}},
\end{equation}
\begin{equation}
\sigma_{1}^2=\frac{1}{\tilde{h}_{i*}'\Pi^{-}\tilde{h}_{i*} + 1/\sigma^{2}_{u}},
\end{equation}
\begin{equation}
\tilde{\epsilon}_{i*}=\tilde{y}_{i*}-\mathbf{\tilde{x}}_{i*}\beta
\end{equation}
The log likelihood function of the model $L$ is obtained by summing the marginal likelihood over $i=1,\ldots,I$
\begin{equation}
L=\Sigma^{I}_{i=1}L_{i}
\end{equation} 
\subsection{Inefficiency and Fixed-Effect Estimation}
The inefficiency index of observation/firm, $i$, during period, $t$, can be estimated as the expection of $u_{it}$ conditional on the model residue, $\tilde{\epsilon}_{i*}~:$
\begin{equation}
\label{eqn:inefind}
E\left(u_{it}|\tilde{\epsilon}_{i*}\right)=h_{it}\left[\mu_{1}+\frac{\phi\left(\frac{\mu_{1}}{\sigma_1}\right)\sigma_{1}}{\Phi\left(\frac{\mu_{1}}{\sigma_1}\right)}\right]
\end{equation}
The fixed-effects, $\alpha_{i}$'s, can be recovered from the estimates of parameters obtained, 
\begin{equation}
\label{eq:fixeff}
\hat{\alpha}_{i}=y_{i*}-\boldsymbol{x}_{i*}\boldsymbol{\hat{\beta}}+\hat{\mu}_{2}\hat{h}_{i*}+\hat{\sigma}_{2}\hat{h}_{i*}\frac{\phi\left(\frac{\hat{\mu}_{2}}{\hat{\sigma}_2}\right)}{\Phi\left(\frac{\hat{\mu}_{2}}{\hat{\sigma}_2}\right),}
\end{equation}
where
\begin{equation}
\hat{\mu}_{2}=\frac{\hat{\mu}\hat{\sigma}^{-2}_{u}-\hat{\sigma}^{-2T}_{v}\sum_{t}\hat{\epsilon}_{it}\hat{h}_{it}}{\hat{\sigma}^{-2T}_{v}\sum_{t}\hat{h}_{it}^{2}+\hat{\sigma}^{-2}_{u}}
\end{equation}
\begin{equation}
\hat{\sigma}^{2}_{2}=\frac{\hat{\sigma}^{2T}_{v}}{\sum_{t}\hat{h}_{it}^{2}+\hat{\sigma}^{2T}_{v}\hat{\sigma}^{-2}_{u}}
\end{equation}
%\subsection{R-Code}
A routine, in the R-statistical language, is written to estimate the maximum likelihood function (Eq. \ref{eqn:loglik}). Additional routines computes inefficiency indices following Eq. \ref{eqn:inefind} and the firm fixed-effects following Eq. \ref{eq:fixeff}. The R-code is tested with STATA procedure and test data obtained from Hun-Jen Wang, as described in detail in \cite{Wang2010}. In addition Monte Carlo simulations are done to test the R-routine. The complete R-code is available freely from the authors on request. 
\bibliographystyle{apalike} 
%\onehalfspacing
%\singlespacing
\bibliography{chap02refs}
%\doublespacing

% chapter02 tables

% Table generated by Excel2LaTeX from sheet 'Sheet2'
%\pagestyle{empty}
\begin{sidewaystable}[htbp]
\renewcommand{\arraystretch}{1.2}  
  \centering  
  %\vspace{-2cm}
  \caption{Variable Definition and Units-I}  
  \label{tab:vardef}
  \scalebox{0.7}{
    \begin{tabular}{lp{0.2\textwidth}p{0.2\textwidth}p{0.4\textwidth}p{0.4\textwidth}}
    \toprule
    \textbf{Variable} & \textbf{Units} & \textbf{Description} & \textbf{Source} \\
    \midrule
    1. & \emph{\textbf{Electricity Output}} & GWhr & Total electricity generated or distributed. Computed by dividing the reported revenue from operations by yearly average region-wise electricity for each type of generating technology. In case of T\&D and vertically integrated utilities the region-wise yearly average retail electricity prices are used. & (a) Company operating revenue reported in annual reports obtained  from CMIE PROWESS.$^{a}$ (b) Electricity retail prices obtained from TEDDY year-book.$^{b}$ \\
    2. & \emph{\textbf{Capital Deployed}} & million Indian Rupees (INR) & Gross fixed assets (real) deployed. Current values deflated by GDP (1999=100). Computed for a period by adding Net-fixed assets of the period with cumulated depreciation till that period.  & (a) Company assets and depreciation reported in annual reports obtained from CMIE PROWESS. (b) GDP obtained from World Bank Development Indicators$^{c}$. \\
    3. & \emph{\textbf{Labor Employed}} & Numbers & Computed by dividing the total reported employee expenditure by the yearly average estimated wages in the power sector in India. & (a) Company total employee expenditure reported in annual reports obtained from CMIE PROWESS. (b) Yearly average wages in power sector estimated by a smaller sample of firm data reporting both number of people employed and the total expenditure on labor. This smaller sample is obtained from CMIE PROWESS and DATASTREAM financial data. \\
    4. & \emph{\textbf{\mbox{Fuel~Consumed}: Coal}} & GWhr energy equivalent of coal used. & Computed by dividing the total reported expenditure on fuel/raw-material by the yearly average estimated purchase price of coal in each region. An average calorific value of 4000KCal/Kg or 4648.9KWhr/metric tonne is assumed for coal. & (a) Company total expenditure on fuel/raw-material reported in annual reports obtained from CMIE PROWESS. (b) Yearly average purchase price of coal in power sector estimated by a smaller sample of firm data reporting both quantity of fuel and the total expenditure on fuel from each region. This smaller sample is obtained from CMIE PROWESS. \\
    5. & \emph{\textbf{\mbox{Fuel~Consumed}: Gas}} & GWhr energy equivalent of natural gas used. & Computed by dividing the total reported expenditure on fuel/raw-material by the yearly average estimated purchase price of gas in each region. An average calorific value of 40 Mjoule/m3 or 11.11KWhr/m3 is assumed. & (a) Company total expenditure on fuel/raw-material reported in annual reports obtained from CMIE PROWESS. (b) Yearly average purchase price of natural gas in power sector estimated by a smaller sample of firm data reporting both quantity of fuel and the total expenditure on fuel from each region. This smaller sample is obtained from CMIE PROWESS. \\
    6. & \emph{\textbf{Electricity Input}} & GWhr & Computed by dividing the total reported expenditure on fuel/electricity purchased by the yearly average sale price to utilities in each region. & (a) Company total expenditure on fuel/electricity purchase reported in annual reports obtained from CMIE PROWESS. (b) Yearly average sale price of electricity by Coal and Gas based generators separately. Estimated by a smaller sample of firm data, from CMIE PROWESS, reporting both quantity of electricity sales and the total revenue from electricity sales from each region.  \\
    \bottomrule
    \multicolumn{4}{r}{\emph{Continued on Table.\ref{tab:vardef01} \ldots}}

    \end{tabular}%
    }  
\end{sidewaystable}%

% Table generated by Excel2LaTeX from sheet 'Sheet2'
\newpage
%\thispagestyle{empty}
%\addtocounter{table}{-1}
\begin{sidewaystable}[htbp]
\renewcommand{\arraystretch}{1.2}  
  \centering  
  %\vspace{-0.5cm}
  \caption{Variable Definition and Units-II}  
   \label{tab:vardef01}
  \scalebox{0.77}{    \begin{tabular}{lp{0.25\textwidth}p{0.17\textwidth}p{0.5\textwidth}p{0.2\textwidth}}
    \multicolumn{4}{l}{\ldots Continued from Table.\ref{tab:vardef}}\\
    \toprule
    & \textbf{Variable} & \textbf{Units} & \textbf{Description} & \textbf{Source} \\
    \midrule
    7. & \emph{\textbf{Coal Price}} & INR per metric tonne & Region-Year average purchase price of coal in power sector estimated by a smaller sample of firm data reporting both quantity of fuel and the total expenditure on fuel from each region.  & The smaller sample of firm data obtained from CMIE PROWESS. \\
    8. & \emph{\textbf{Gas Price}} & INR per cubic meter & Region-Year average purchase price of natural in power sector estimated by a smaller sample of firm data reporting both quantity of fuel and the total expenditure on fuel from each region. & The smaller sample of firm data obtained from CMIE PROWESS. \\
    9. & \emph{\textbf{\mbox{Sale Price of Electricity by} \mbox{Coal~Based} Generators}} & INR per KWhr & Region-Year average sale price of electricity by Coal based generators. Estimated by a smaller sample of firm data reporting both quantity of electricity sales and the total revenue from electricity sales from each region.  & The smaller sample of firm data obtained from CMIE PROWESS. \\
    10. & \emph{\textbf{\mbox{Sale Price of Electricity by} \mbox{Gas~Based Generators}}} & INR per KWhr & Region-Year average sale price of electricity by Gas based generators. Estimated by a smaller sample of firm data reporting both quantity of electricity sales and the total revenue from electricity sales from each region.  & The smaller sample of firm data obtained from CMIE PROWESS. \\
    11. & \emph{\textbf{Retail Price of Electricity}} & INR per KWhr & State-Year average sale price of electricity by utilities.  & Electricity retail prices obtained from TEDDY year-book \\
    12. & \emph{\textbf{Price of Capital}} & Percentage & Computed as\ldots (a) Price of Capital = Expense of Capital/Gross Fixed Assets. (b) Expense of Capital= Interest Share of Capital + Depreciation. (c) Interest Share of Capital=(Annual Interest on Long-Term Debt)*(Fixed Assets)/(Long Term Debt) & Company financial data in annual reports obtained  from CMIE PROWESS.  \\
    13. & \emph{\textbf{Vintage}} & Year & The year of incorporation of the company is taken as a proxy for the approximate vintage of the firm's productive assets. & Company annual reports. \\
    14. & \emph{\textbf{Time Since Un-bundling}} & Year & Years past since the vertically integrated state electricity boards (SEBs) were unbundled and separated as generators and distribution utilities in the respective states. & TEDDY year-book \\
%    15. & \emph{\textbf{Index of Financial Sustainability of Power Utilities in the State}} & Index Number & Measure of the sustainability of revenue model of the state power sector. Score of 0-to-40, higher score indicating more sustainable revenue model. & Ministry of Power, Government of India$^{d}$ \\
    16. & \emph{\textbf{\mbox{Index of Competition in} \mbox{Power Sector in the State}}} & Index Number &Measure of the competitiveness of the state power sector. Score of 0-to-40, higher score indicating more sustainable revenue model & Ministry of Power, Government of India \\
    %\midrule
    \bottomrule
    \multicolumn{5}{l}{(a) CMIE PROWESS: The PROWESS database of Indian companies details maintained by the Center for Monitoring of Indian Economy, Mumbai India.}\\
    \multicolumn{5}{l}{(b) TEDDY: TERI Energy Data Directory and Yearbook, annual publication of The Energy and Resource Institute, New Delhi India.  }\\
    \multicolumn{5}{l}{(c) World Bank Development Indicators: maintained at http://data.worldbank.org/data-catalog/world-development-indicators}\\
    \multicolumn{5}{l}{(d) Ministry of Power Report: http://www.powermin.nic.in/ indian\_electricity\_scenario/pdf/Final\_Report\_Rating.pdf}
    %\bottomrule
    \end{tabular}%
} %
\end{sidewaystable}%

%\hspace{-2cm}
\begin{sidewaystable}[htbp]%\footnotesize
%\vspace{-1cm}
%\hspace{2cm}
\renewcommand{\arraystretch}{1.2}
\centering
\caption{Description of Sample}
 \label{tab:02sample}
\scalebox{1}{
\begin{tabular}{lcccccc.c.}
    \toprule     
     \multicolumn{1}{l}{} & \multicolumn{3}{c}{Generation}\\
     \cline{2-4}
     \multicolumn{1}{l}{} & \multicolumn{2}{c}{Fossil Fuels} & \multicolumn{1}{c}{Mixed} 
& \multicolumn{2}{c}{Transmission \& Distribution}\\
     \cline{2-3}     
     \cline{5-6}
     \multicolumn{1}{l}{Year} & \multicolumn{1}{c}{Coal} & \multicolumn{1}{c}{Gas} 
& \multicolumn{1}{c}{} & \multicolumn{1}{p{2cm}}{Distribution Utilities}
& \multicolumn{1}{p{1cm}}{Vertically Integrated} & \multicolumn{1}{p{2cm}}{Total Generation (GWhr)} 
& \multicolumn{1}{p{2cm}}{Generation Sample Coverage} & \multicolumn{1}{p{2cm}}{Total Consumption (GWhr)}
& \multicolumn{1}{p{2cm}}{Distribution Sample Coverage}\\
     \cline{1-10}
2000 & 8 & 4 & 3 & 5 & 3 & 501,204 & 33.9\% & 316,600 & 27.3\% \\
2001 & 8 & 5 & 3 & 9 & 7 & 517,439 & 31.3\% & 322,459 & 70.0\% \\
2002 & 16 & 9 & 5 & 11 & 7 & 532,693 & 45.6\% & 339,598 & 57.9\% \\
2003 & 16 & 11 & 5 & 17 & 6 & 565,102 & 50.8\% & 360,937 & 62.0\% \\
2004 & 15 & 13 & 4 & 23 & 6 & 594,456 & 42.7\% & 386,134 & 58.5\% \\
2005 & 14 & 15 & 6 & 25 & 6 & 623,820 & 47.5\% & 411,887 & 67.4\% \\
2006 & 14 & 18 & 2 & 25 & 4 & 670,654 & 48.6\% & 455,748 & 61.4\% \\
2007 & 13 & 16 & 5 & 30 & 4 & 722,626 & 48.5\% & 501,977 & 70.5\% \\
2008 & 12 & 17 & 6 & 25 & 4 & 741,167 & 48.4\% & 527,564 & 63.0\% \\
2009 & 15 & 17 & 3 & 23 & 4 & 790,766 & 51.9\% & 553,151 & 50.1\% \\
\midrule
Firms=98 & 21 & 19 & 11 & 38 & 9 & 6,259,927 & 45.7\% & 4,176,055 & 59.4\% \\
Firm-Years=542 & 131 & 125 & 42 & 193 & 51 \\
\bottomrule
    \end{tabular}
    }
\end{sidewaystable}

\begin{table}[htbp]%\footnotesize
  \renewcommand{\arraystretch}{1.5}
  \centering
  \caption{Summary Statistics}
  \label{tab:02sum1}
  \scalebox{0.85}{
  \begin{threeparttable}
  \begin{tabular}{l.....}
  \toprule     
  \multicolumn{1}{l}{Variable} & \multicolumn{1}{c}{Mean} & \multicolumn{1}{c}{S.d.} & \multicolumn{1}{c}{Min} & \multicolumn{1}{c}{Max} & \multicolumn{1}{c}{N} \\
  \cline{1-6}   
Electricity Output\tnote{a} & 9,848.67 & 20,111.82 & 16.29 & 207,351.79 & 542\\
Capital Deployed\tnote{b} & 38,992.40 & 72,951.23 & 160.60 & 668,474.20 & 542\\
Labour\tnote{c} & 4,800.83 & 9,006.34 & 2.25 & 77,528.54 & 542\\
Fuel:Coal\tnote{d} & 63,731.08 & 141,227.98 & 117.04 & 738,319.02 & 131\\
Fuel:Gas\tnote{e} & 8,055.95 & 18,477.04 & 56.94 & 102,879.34 & 125\\
Electricity Input\tnote{f} & 11,858.73 & 14,095.34 & 0.55 & 89,783.45 & 193\\
Coal Price\tnote{g} & 1,533.57 & 246.48 & 1,227.40 & 1,839.40 & 542\\
Gas Price\tnote{h} & 5.21 & 0.88 & 4.00 & 6.60 & 542\\
Captial Price\tnote{i} & 8.68\% & 5.43\% & 0.10\% & 42.00\% & 542\\
Coal Gen. Price\tnote{j} & 2.10 & 0.57 & 0.64 & 3.26 & 131\\
Gas Gen. Price\tnote{k} & 3.05 & 1.51 & 1.14 & 9.41 & 125\\
Retail Electricity Price\tnote{l} & 2.11 & 0.76 & 0.48 & 4.05 & 193\\
Year of Incorporation\tnote{m} & 1,987.15 & 22.35 & 1,913.00 & 2,008.00 & 542\\
Time Since Unbundled\tnote{n} & 2.63 & 4.72 & -9 & 13 & 542\\
Competition Index\tnote{p} & 23.49 & 7.45 & 0.00 & 40.00 & 542\\
\bottomrule   
 \end{tabular}
 \begin{tablenotes}
 \smallskip
       \item \emph{Variable Definition and Units}
       \smallskip 
       \item[a] Electricity generated or distributed in GWhr. Computed by dividing reported revenue from operations by yearly average regional electricity prices for each generating technology. In case of T\&D and vertically integrated companies the state-wise yearly average retail electricity prices are used. 
       \item[b] Real gross fixed assets deployed in million Indian Rupees, deflated by GDP (1999=100)
       \item[c] No. of employees. Computed by dividing the total reported employee expenditure by the yearly average estimated wages in the power sector.
       \item[d] GWhr equivalent of coal used. Computed by dividing the reported fuel expenditure by the yearly average purchase price of coal obtained from a smaller sample of firms reporting this information. An average calorific value of 4000KCal/Kg or 4648.9KWhr/metric tonne is assumed for coal.
       \item[e] GWhr equivalent of gas used. Computed by dividing the reported fuel expenditure by the yearly average purchase price of gas obtained from a smaller sample of firms reporting this information. An average calorific value of 40 Mjoule/m3 or 11.11KWhr/m3 is assumed.
       \item[f] Electricity purchased in GWhr. Computed by dividing reported expenditure on fuel by yearly average region-wise electricity sale price of generators to distribution utilities.
       \item[g] INR per metric tonne. Region-Year average purchase price of coal used in the power sector.
       \item[h] INR per cubic meter. Region-Year average purchase price of natural in power sector.
       \item[i] Percentage, computed as$:$ Price of Capital = Expense of Capital/Gross Fixed Assets.
       \item[j] INR per KWhr. Region-Year average sale price of electricity by Coal based generators.
       \item[k] INR per KWhr. Region-Year average sale price of electricity by Gas based generators.
       \item[l] State-Year average sale price of electricity by utilities.
       \item[m] Year of incorporation of the firm. used as proxy for asset vintage.
       \item[n] Time in years, since the home State power sector is unbundled.
       \item[p] Index of competitiveness of power sector in the State.(0=low to 40=high)
 \end{tablenotes}
 \end{threeparttable}
 }
 \end{table}

\begin{table}[htbp]%\footnotesize
\begin{adjustwidth}{-1.5cm}{-1cm}
\renewcommand{\arraystretch}{1.5}
\centering
\caption{Variable Statistics, Mean (s.d), by Technology}
\label{tab:02sum2}
\scalebox{0.80}{
\begin{threeparttable}
\begin{tabular}{l.....}
    \toprule     
     \multicolumn{1}{l}{} & \multicolumn{3}{c}{Generation}\\
     \cline{2-4}
     \multicolumn{1}{l}{} & \multicolumn{2}{c}{Fossil Fuels} & \multicolumn{1}{c}{Mixed} 
& \multicolumn{2}{c}{Transmission \& Distribution}\\
     \cline{2-3}     
     \cline{5-6}
     \multicolumn{1}{l}{Year} & \multicolumn{1}{c}{Coal} & \multicolumn{1}{c}{Gas} 
& \multicolumn{1}{c}{} & \multicolumn{1}{p{1.75cm}}{Distribution Utilities}
& \multicolumn{1}{p{1.75cm}}{Vertically Integrated}\\
     \cline{1-6}
Electricity Output\tnote{a} & 17,970.34 & 2,054.39 & 5,898.20 & 8,560.41 & 16,219.20 \\
 & (35,971.94) & (3,614.19) & (6,580.76) & (8,459.28) & (16,848.78) \\
Capital Deployed\tnote{b} & 46,060.92 & 6,648.24 & 11,926.62 & 14,266.73 & 51,048.37 \\
 & (74,565.80) & (7,760.42) & (17,180.15) & (13,024.37) & (53,554.72) \\
Labour\tnote{c} & 5,963.04 & 329.26 & 2,233.28 & 4,759.15 & 15,047.49 \\
 & (9,457.96) & (963.34) & (2,783.19) & (5,595.16) & (18,709.51) \\
Fuel:Coal\tnote{d} & 63,731.08 &  &  &  &  \\
 & (141,227.98) &  &  &  &  \\
Fuel:Gas\tnote{e} &  & 8,055.95 &  &  & \\
 &  & (18,477.04) &  &  &  \\
Electricity Input\tnote{f} &  &  &  & 11,858.73 & 122,108.12 \\
 &  &  &  & (14,095.34) & (132,853.61) \\
\midrule
(Firms:98) & 21 & 19 & 11 & 38 & 9 \\
(Firm-Years:542) & 131 & 125 & 42 & 193 & 51 \\
\bottomrule
    \end{tabular}
        \begin{tablenotes}
       \smallskip
       \item \emph{Variable Definition and Units}
       \smallskip 
       \item[a] Electricity generated or distributed in GWhr. Computed by dividing reported revenue from operations by yearly average regional electricity prices for each generating technology. In case of T\&D and vertically integrated companies the state-wise yearly average retail electricity prices are used. 
       \item[b] Real gross fixed assets deployed in million Indian Rupees, deflated by GDP (1999=100)
       \item[c] No. of employees. Computed by dividing the total reported employee expenditure by the yearly average estimated wages in the power sector.
       \item[d] GWhr equivalent of coal used. Computed by dividing the reported fuel expenditure by the yearly average purchase price of coal obtained from a smaller sample of firms reporting this information. An average calorific value of 4000KCal/Kg or 4648.9KWhr/metric tonne is assumed for coal.
       \item[e] GWhr equivalent of gas used. Computed by dividing the reported fuel expenditure by the yearly average purchase price of gas obtained from a smaller sample of firms reporting this information. An average calorific value of 40 Mjoule/m3 or 11.11KWhr/m3 is assumed.
       \item[f] Electricity purchased in GWhr. Computed by dividing reported expenditure on fuel by yearly average region-wise electricity sale price of generators to distribution utilities.       
     \end{tablenotes}
  \end{threeparttable}
    }  
\end{adjustwidth}
\end{table}

  \begin{table}[htbp]%\footnotesize
  \begin{adjustwidth}{-0cm}{0cm}  
  \vspace{-0.75cm}
  \renewcommand{\arraystretch}{1.2}
  \centering
  \caption{Maximum Likelihood Estimates of the Translog Production Model Parameters}
  \label{tab:model1}
  \scalebox{0.72}{
  \begin{tabular}{llrrrrr}
  \toprule     
  \multicolumn{1}{l}{\textbf{Variable}} &\multicolumn{1}{l}{\textbf{Par.}}& \multicolumn{1}{c}{\textbf{Coal}} & \multicolumn{1}{c}{\textbf{Gas}} & \multicolumn{1}{c}{\textbf{Mixed}} 
& \multicolumn{1}{c}{\textbf{TnD}} & \multicolumn{1}{c}{\textbf{Integr.}} \\
  \cline{1-7}
    
$ln(K)$ & $\beta_{K}$ & -0.412  & 0.205  & 1.686$\dag$ & -2.138  & -0.284* \\
  & & (0.550) & (1.052) & (1.314) & (3.349) & (0.169) \\
$ln(L)$ & $\beta_{L}$ & 1.550*** & 0.864* & 0.210  & -0.070  & -0.720*** \\
  & & (0.364) & (0.399) & (1.083) & (0.867) & (0.048) \\
$ln(F)$ & $\beta_{F}$  & 0.340  & 0.654$\dag$ & 1.924*** & 0.767* & 0.286  \\
  & & (0.287) & (0.464) & (0.520) & (0.363) & (0.349) \\
$\frac{1}{2}ln(K)ln(K)$  & $\beta_{KK}$ & 0.180* & 0.256* & 0.074  & 0.212  & 0.122$\dag$ \\
  & & (0.088) & (0.145) & (0.264) & (0.340) & (0.084) \\
$\frac{1}{2}ln(L)ln(L)$ & $\beta_{LL}$  & -0.113** & 0.119* & 0.330** & -0.111  & 0.361** \\
  & & (0.044) & (0.054) & (0.118) & (0.134) & (0.140) \\
$\frac{1}{2}ln(F)ln(F)$ & $\beta_{FF}$  & 0.050* & 0.224*** & 0.436*** & 0.070*** & 0.022  \\
  & & (0.023) & (0.035) & (0.075) & (0.013) & (0.085) \\
$\frac{1}{2}ln(K)ln(L)$ & $\beta_{KL}$  & -0.062  & -0.201$\dag$ & 0.232  & 0.222$\dag$ & -0.331* \\
  & & (0.105) & (0.153) & (0.308) & (0.165) & (0.151) \\
$\frac{1}{2}ln(K)ln(F)$ & $\beta_{KF}$  & -0.079  & -0.380*** & -0.658*** & -0.206** & 0.112  \\
  & & (0.065) & (0.099) & (0.192) & (0.081) & (0.177) \\
$\frac{1}{2}ln(L)ln(F)$ & $\beta_{LF}$  & -0.049  & -0.076  & -0.668*** & 0.062$\dag$ & -0.179$\dag$ \\
  & & (0.041) & (0.095) & (0.193) & (0.039) & (0.120) \\
$Time$ & $\beta_{t}$  & -0.043  & 0.247$\dag$ & 0.114$\dag$ & 0.065  & 0.253* \\
  & & (0.072) & (0.184) & (0.088) & (0.221) & (0.111) \\
$\frac{1}{2}Time^2$ & $\beta_{tt}$  & 0.026** & -0.020* & -0.003  & -0.015* & -0.046*** \\
  & & (0.009) & (0.010) & (0.011) & (0.008) & (0.012) \\
$ln(K)Time$ & $\beta_{Kt}$  & -0.035*** & 0.035  & 0.062* & 0.038* & -0.039** \\
  & & (0.010) & (0.030) & (0.033) & (0.022) & (0.015) \\
$ln(L)Time$ & $\beta_{Lt}$  & 0.028*** & -0.022  & 0.021  & -0.014  & 0.083*** \\
  & & (0.007) & (0.018) & (0.041) & (0.020) & (0.013) \\
$ln(F)Time$ & $\beta_{Ft}$  & 0.009* & -0.031* & -0.077** & -0.012* & -0.023  \\
  & & (0.005) & (0.016) & (0.026) & (0.006) & (0.027) \\
\midrule
\multicolumn{5}{l}{\textbf{Exogenous explanatory variables} } \\
\midrule   
$Time$ & $\delta_{t}$  & -0.192  & 0.321$\dag$ & -0.944  & 0.924*** & 0.080  \\
  & & (0.211) & (0.219) & (1.107) & (0.290) & (0.085) \\
$Time^2$ & $\delta_{tt}$  & 0.018  & -0.023$\dag$ & -1.318  & -0.054** & -0.007  \\
  & & (0.019) & (0.016) & (3.154) & (0.018) & (0.008) \\
$Asset~Vintage$ & $\delta_{V}$  & -0.239  & 0.217  & 0.506  & -1.496* & 0.126  \\
  & & (0.191) & (0.670) & (1.102) & (0.883) & (0.139) \\
$Owner:Central~Govt.$ & $\delta_{CG}$  & -0.835*** & -- & -- & -0.461  & -- \\
  & & (0.076) & -- & -- & (1.601) & -- \\
$Owner:State~Govt.$ & $\delta_{SG}$  & -0.045  & 0.193  & -0.106*** & 0.043  & 1.033*** \\
  & & (0.413) & (0.753) & (0.000) & (0.369) & (0.050) \\
$Unbundled$ & $\delta_{Udl}$  & 0.081  & 0.034  & -0.079  & -1.803  & --  \\
  & & (0.100) & (0.090) & (0.590) & (1.441) & -- \\
$Competition$ & $\delta_{Cmp}$  & 0.025  & -0.281$\dag$ & -0.483  & -0.086  & -0.045  \\
  & & (0.140) & (0.210) & (1.282) & (0.121) & (0.315) \\
\midrule
\multicolumn{5}{l}{\textbf{Inefficiency}} \\
\midrule   
& $ln(\sigma_{u})$  & 2.585*** & -0.297*** & 0.174  & -1.052$\dag$ & 0.087*** \\
  & & (0.013) & (0.066) & (0.165) & (0.722) & (0.010) \\
 & $ln(\sigma_{v})$  & -4.422*** & -3.967*** & -4.980*** & -4.028*** & -6.420*** \\
  & & (0.148) & (0.145) & (0.258) & (0.128) & (0.251) \\
\midrule
$Log~Likelihood$ &  & 72.676  & 54.149  & 32.997  & 69.358  & 62.061  \\
\bottomrule 
\multicolumn{7}{l}{Standard errors (in parenthesis) computed using delta method.}\\
\multicolumn{7}{l}{Significance denoted by $\dag:~p<0.1,~*:~p<0.05,~**:~p<0.01,~***:~p<0.001$} 
 \end{tabular}
  }    
\end{adjustwidth} 
\end{table}

  \begin{sidewaystable}[htbp]%\footnotesize
  \renewcommand{\arraystretch}{1.3}
  \centering
  \caption{Power Sector TFP Changes and Decomposition of TFP }
  \label{tab:prod}
  \scalebox{0.76}{
  \begin{tabular}{l|.rrrrr|rrrrrr|rrrrrr}
  \toprule     
  \multicolumn{1}{l}{Year} & \multicolumn{6}{c}{Generator:Coal} 
& \multicolumn{6}{c}{Generator:Gas} & \multicolumn{6}{c}{Generator:Mixed} 
\\
  \midrule  & \multicolumn{1}{c}{$\dot{TFP}$} & $\Delta T$ & $\Delta TE$ & $\Psi$ & $\Omega$ & $\Gamma$ & $\dot{TFP}$ & $\Delta T$ & $\Delta TE$ & $\Psi$ & $\Omega$ & $\Gamma$ & $\dot{TFP}$ & $\Delta T$ & $\Delta TE$ & $\Psi$ & $\Omega$ & $\Gamma$ \\ \midrule
2000-01  &   0.212  &  -0.034  &   0.158  &   0.095  &  -0.008  &  0.554  &   0.156  &   0.148  &  -0.076  &   0.092  &  -0.008  &  -0.258  &   0.336  &  0.057  &  0.093  &   0.193  &  -0.007  &  -18.704 \\
2001-02  &   0.207  &  -0.018  &   0.125  &   0.006  &   0.094  &  1.714  &   0.016  &   0.146  &  -0.093  &   0.015  &  -0.052  &   0.916  &   0.205  &  0.030  &  0.002  &   0.182  &  -0.008  &  -19.211 \\
2002-03  &   0.068  &   0.015  &   0.084  &  -0.023  &  -0.007  &  1.550  &  -0.017  &   0.122  &  -0.076  &  -0.066  &   0.003  &   0.794  &  -0.950  &  0.129  &  0.000  &  -1.098  &   0.018  &  -13.252 \\
2003-04  &   0.063  &   0.039  &   0.043  &  -0.008  &  -0.012  &  1.781  &   0.054  &   0.096  &  -0.058  &   0.040  &  -0.025  &   0.632  &  -0.029  &  0.100  &  0.000  &  -0.126  &  -0.002  &  -14.647 \\
2004-05  &  -0.115  &   0.065  &  -0.003  &  -0.151  &  -0.026  &  1.831  &  -0.392  &   0.073  &  -0.044  &  -0.383  &  -0.038  &   0.812  &  -1.430  &  0.188  &  0.000  &  -1.624  &   0.006  &   -7.765 \\
2005-06  &  -0.001  &   0.100  &  -0.024  &  -0.058  &  -0.019  &  1.360  &   0.090  &   0.036  &  -0.022  &   0.103  &  -0.027  &   0.546  &   2.766  &  0.006  &  0.000  &   2.787  &  -0.026  &  -25.100 \\
2006-07  &   0.190  &   0.136  &  -0.071  &   0.124  &   0.001  &  0.268  &   0.013  &   0.004  &  -0.001  &   0.030  &  -0.021  &   0.334  &   0.285  &  0.032  &  0.000  &   0.256  &  -0.003  &  -18.676 \\
2007-08  &   0.244  &   0.161  &  -0.120  &   0.212  &  -0.008  &  0.662  &  -0.001  &  -0.014  &   0.015  &   0.022  &  -0.024  &   0.718  &  -0.434  &  0.037  &  0.000  &  -0.470  &  -0.001  &  -18.774 \\
2008-09  &   0.091  &   0.188  &  -0.158  &   0.474  &  -0.412  &  0.469  &  -0.112  &  -0.034  &   0.045  &  -0.112  &  -0.011  &   0.548  &  -2.689  &  0.005  &  0.000  &  -2.699  &   0.005  &  -26.641 \\
\midrule
$mean_{2000-04}^a$  &   0.138  &  0.001  &   0.103  &   0.018  &   0.017  &    1.400  &   0.052  &  0.128  &  -0.076  &   0.021  &  -0.021  &    0.521  &  -0.109  &  0.079  &   0.024  &  -0.212  &   0.000  &  -16.454 \\
$mean_{2004-09}$  &   0.082  &   0.130  &  -0.075  &   0.120  &  -0.093  &    0.918  &  -0.080  &   0.013  &  -0.001  &  -0.068  &  -0.024  &    0.592  &  -0.300  &   0.054  &   0.000  &  -0.350  &  -0.004  &  -19.391 \\
\midrule
  & \multicolumn{6}{c}{Distribution} & \multicolumn{6}{c}{Integrated}  & \multicolumn{6}{c}{Power Sector All}\\
  \midrule  & \multicolumn{1}{r}{$\dot{TFP}$} & $\Delta T$ & $\Delta TE$ & $\Psi$ & $\Omega$ & $\Gamma$ & $\dot{TFP}$ & $\Delta T$ & $\Delta TE$ & $\Psi$ & $\Omega$ & $\Gamma$ & $\dot{TFP}$ & $\Delta T$ & $\Delta TE$ & $\Psi$ & $\Omega$ & $\Gamma$ \\ \midrule
2000-01  &   1.761  &  0.155  &  -0.032  &   1.632  &   0.005  &  19.892  &   0.167  &   0.227  &  -0.130  &   0.068  &   0.002  &  6.713  &  0.606  &  0.084  &  0.034  &  0.490  &  -0.003  &  3.343 \\
2001-02  &   0.487  &  0.146  &  -0.053  &   0.392  &   0.001  &  20.771  &   0.202  &   0.214  &  -0.169  &   0.146  &   0.011  &  8.586  &  0.239  &  0.105  &  -0.031  &  0.147  &  0.018  &  5.744 \\
2002-03  &   0.951  &  0.133  &  -0.086  &   0.900  &   0.005  &  19.681  &   0.547  &   0.147  &  -0.127  &   0.506  &   0.021  &  8.590  &  0.213  &  0.089  &  -0.019  &  0.140  &  0.003  &  4.970 \\
2003-04  &   0.202  &  0.119  &  -0.121  &   0.205  &  -0.001  &  19.304  &  -0.358  &   0.102  &  -0.103  &  -0.345  &  -0.012  &  8.441  &  0.045  &  0.088  &  -0.049  &  0.017  &  -0.011  &  6.270 \\
2004-05  &   0.525  &  0.101  &  -0.124  &   0.544  &   0.004  &  18.444  &  -0.527  &   0.041  &  -0.027  &  -0.523  &  -0.017  &  7.663  &  -0.032  &  0.083  &  -0.060  &  -0.039  &  -0.016  &  7.923 \\
2005-06  &  -0.359  &  0.101  &  -0.130  &  -0.329  &  -0.002  &  20.401  &  -0.421  &  -0.019  &   0.009  &  -0.404  &  -0.007  &  7.871  &  -0.056  &  0.072  &  -0.063  &  -0.052  &  -0.013  &  8.311 \\
2006-07  &  -0.054  &  0.094  &  -0.100  &  -0.045  &  -0.003  &  20.962  &  -0.432  &  -0.060  &   0.047  &  -0.414  &  -0.005  &  7.868  &  0.012  &  0.066  &  -0.053  &  0.006  &  -0.007  &  7.924 \\
2007-08  &  -0.069  &  0.066  &  -0.059  &  -0.074  &  -0.001  &  20.343  &  -0.067  &  -0.102  &   0.058  &  -0.021  &  -0.002  &  8.003  &  -0.019  &  0.049  &  -0.038  &  -0.021  &  -0.009  &  7.202 \\
2008-09  &   0.316  &  0.058  &   0.007  &   0.249  &   0.001  &  20.339  &  -0.314  &  -0.141  &   0.078  &  -0.247  &  -0.004  &  8.041  &  -0.049  &  0.042  &  -0.011  &  0.007  &  -0.086  &  7.373 \\
\midrule
$mean_{2000-04}$  &   0.850  &  0.138  &  -0.073  &   0.782  &   0.003  &   19.912  &   0.139  &  0.172  &  -0.132  &   0.094  &   0.006  &    8.083  &  0.276  &  0.092  &  -0.016  &  0.198  &  0.002  &  5.082 \\
$mean_{2004-09}$ &   0.072  &   0.084  &  -0.081  &   0.069  &   0.000  &   20.098  &  -0.352  &  -0.056  &   0.033  &  -0.322  &  -0.007  &    7.889  &  -0.016  &  0.067  &  -0.046  &  -0.014  &  -0.024  &  7.501 \\
\bottomrule  
\multicolumn{15}{l}{a: Mean year-on-year changes.}\\ 
\multicolumn{15}{l}{$\Delta T:$ Technology change, $\Delta TE:$ Technical efficiency change, $\Psi:$ Scale effect, $\Omega:$ Price effect, $\Gamma:$ Returns to scale} 
 \end{tabular}
  }  
  \end{sidewaystable}
%\pagestyle{plain}  

%chapter02 figures

\begin{figure}[h]
	\centering
	\caption{Power Sector Technical Efficiency Distribution}
	\label{fig:EffDistr}
		\includegraphics[width=1.00\textwidth]{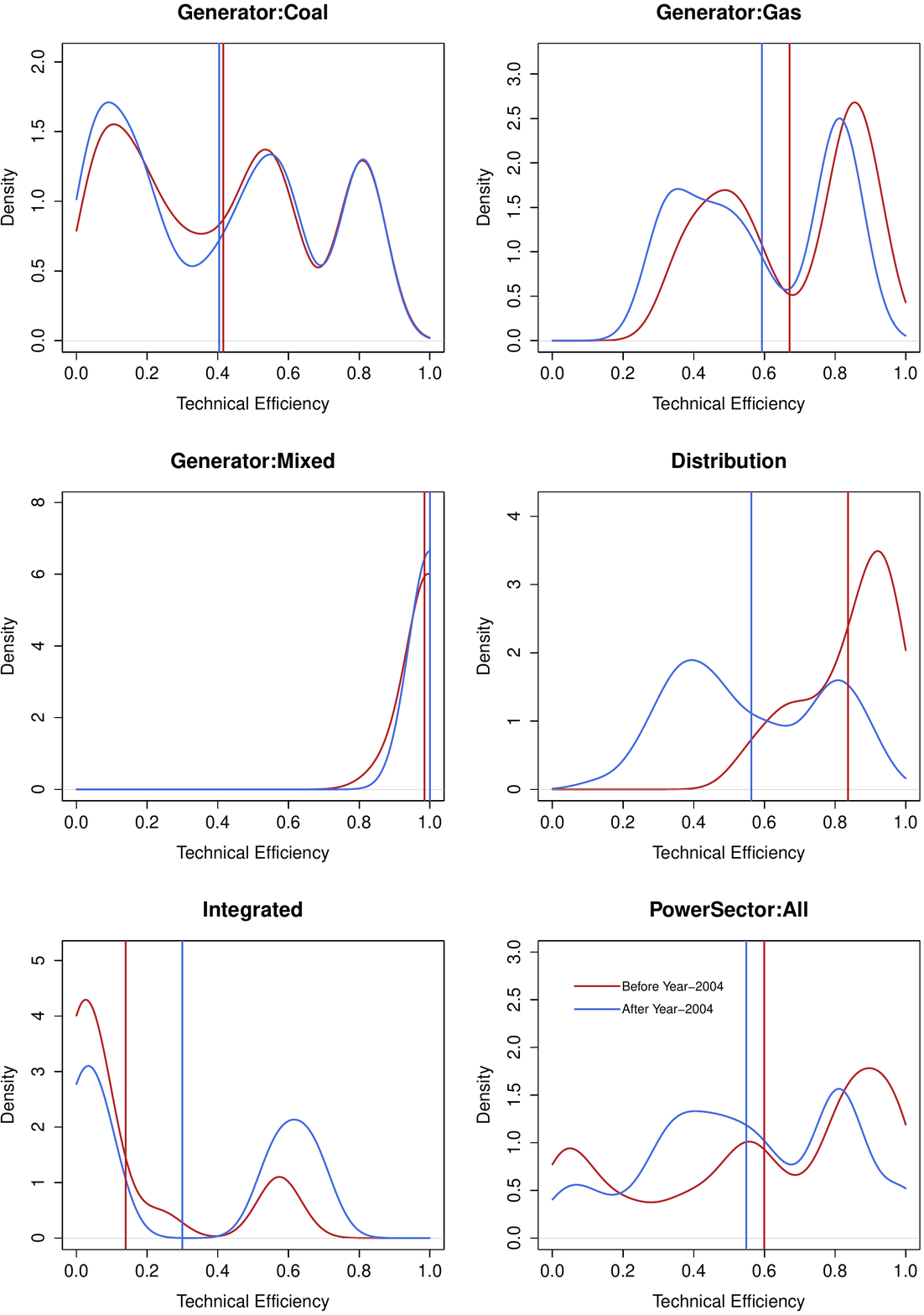}	
\end{figure}

\begin{figure}[h]
\centering
\caption{TFP Change in Power Sector}
	\label{fig:TFPChange}
	\includegraphics[width=1.00\textwidth]{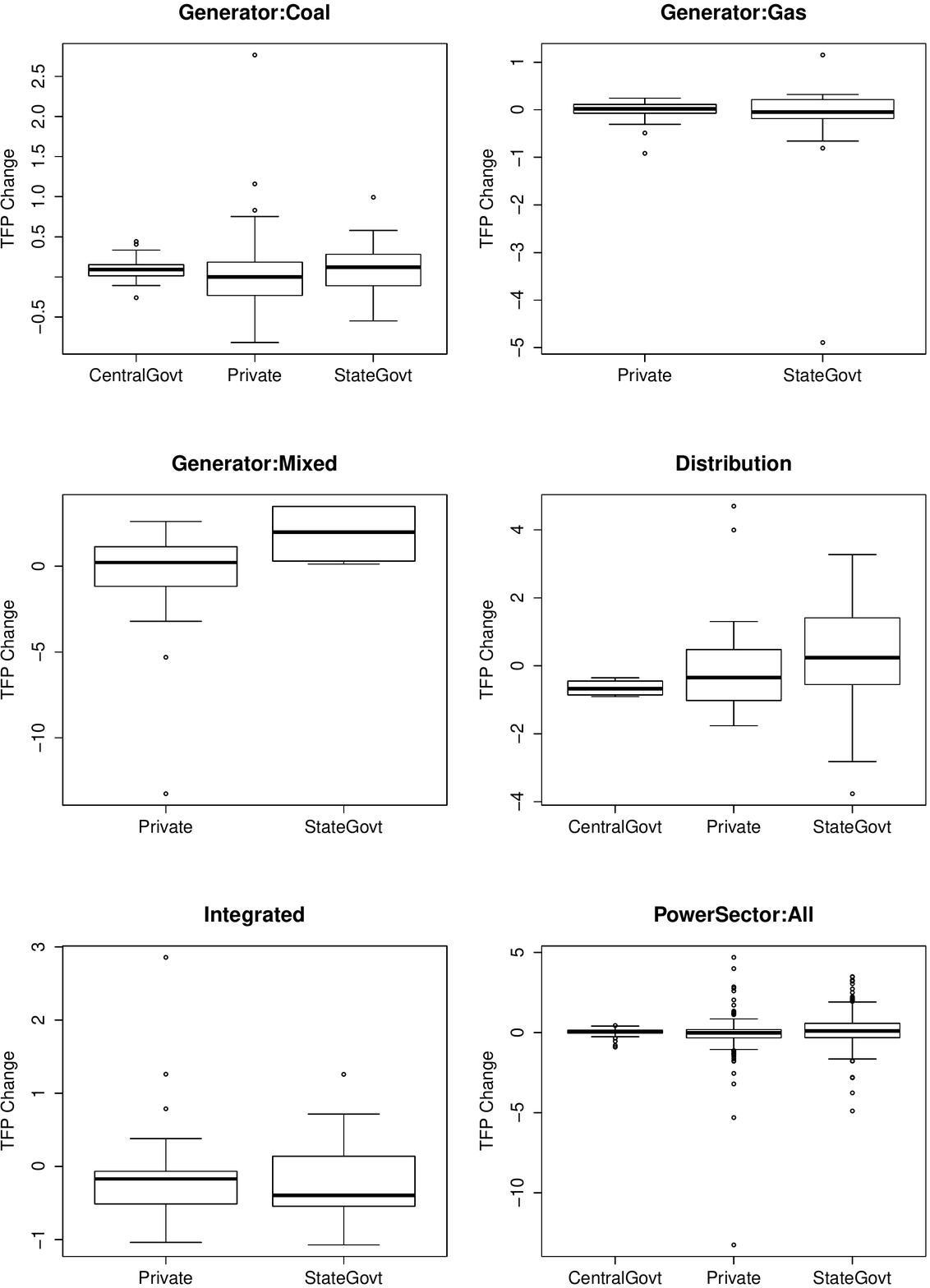}	
\end{figure}

\begin{figure}[h]
	\centering
	\caption{Technical Change in Power Sector}
	\label{fig:TechChange}
		\includegraphics[width=1.00\textwidth]{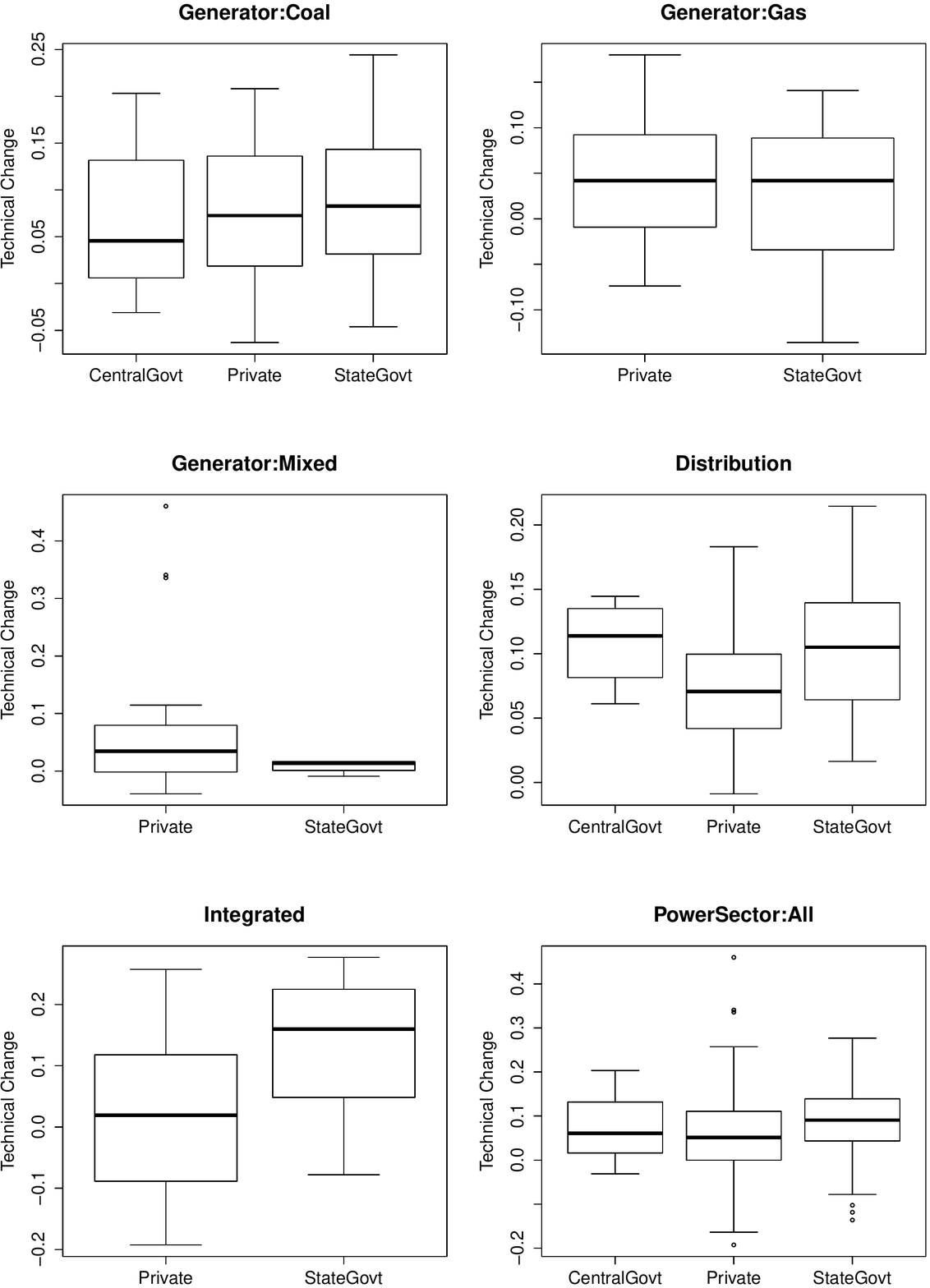}	
\end{figure}

\begin{figure}[h]
\centering
\caption{Efficiency Change in Power Sector}
	\label{fig:EffChange}
\includegraphics[width=1.00\textwidth]{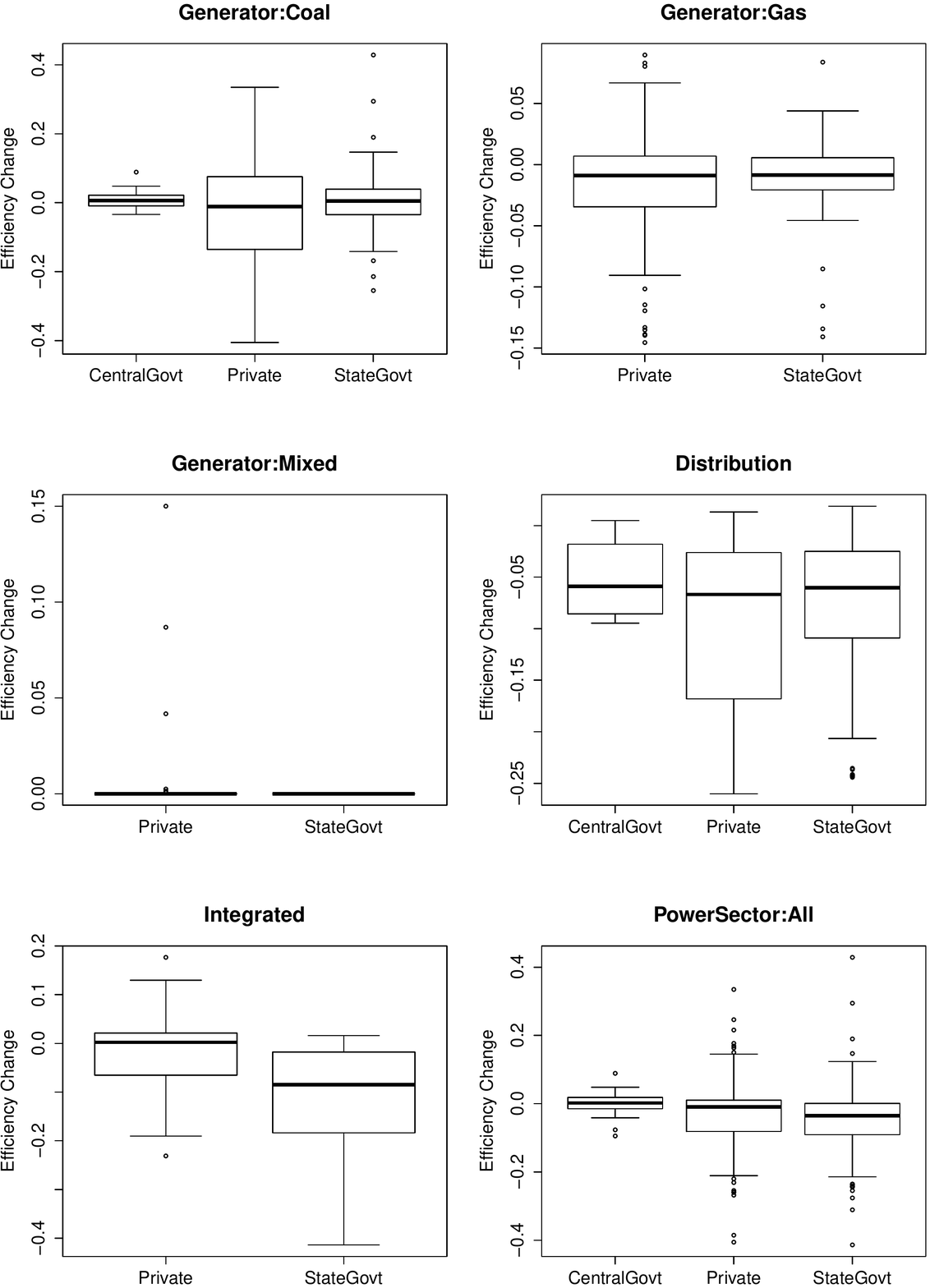}\\		
\end{figure}

\begin{figure}[h]
\centering
\caption{Scale Effect on TFP Change in Power Sector}
	\label{fig:ScaleChange}
	\includegraphics[width=1.00\textwidth]{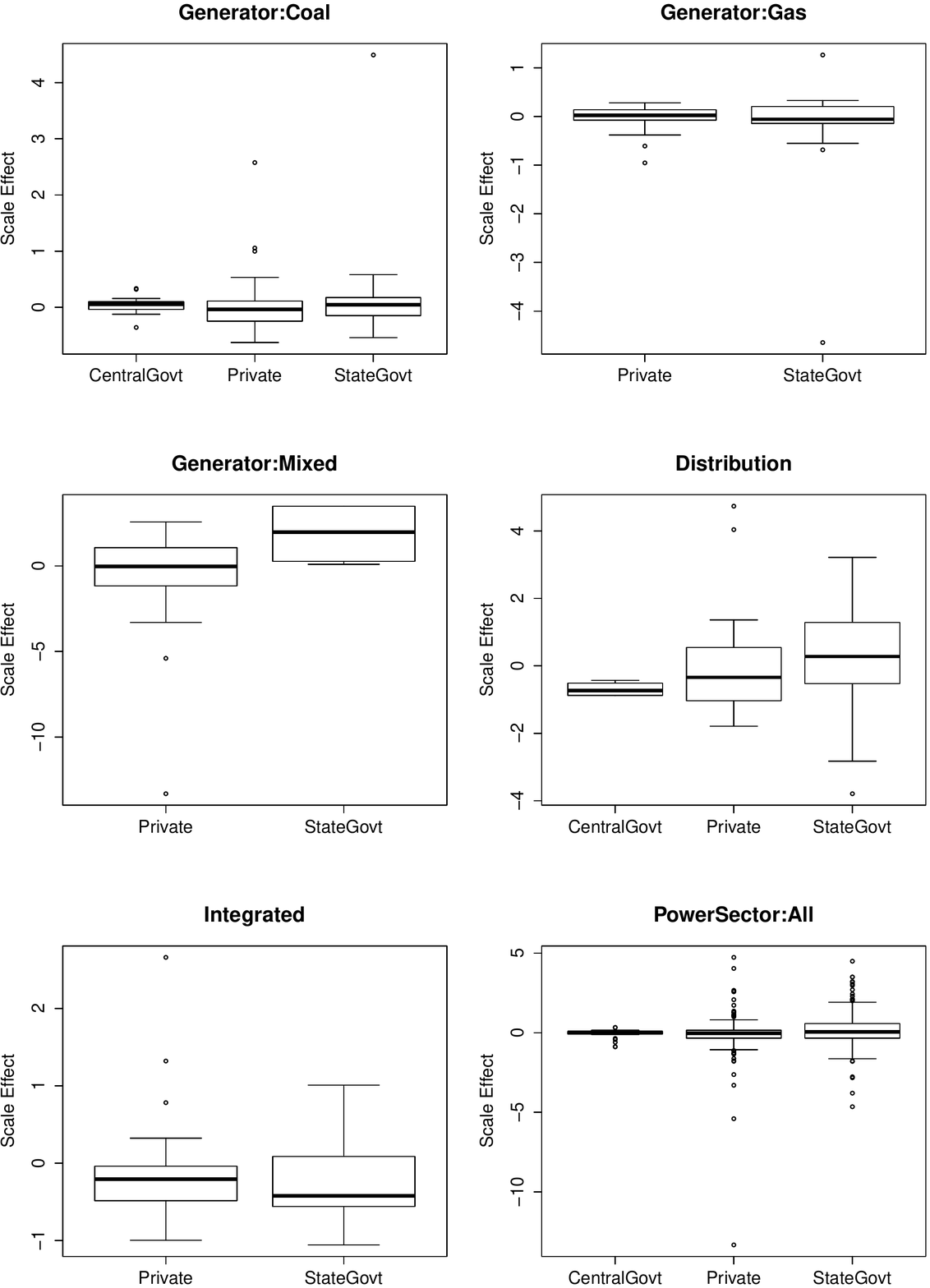}	
\end{figure}

\begin{figure}[h]
\centering
\caption{Price Effect on TFP Change in Power Sector}
	\label{fig:PriceChange}
	\includegraphics[width=1.00\textwidth]{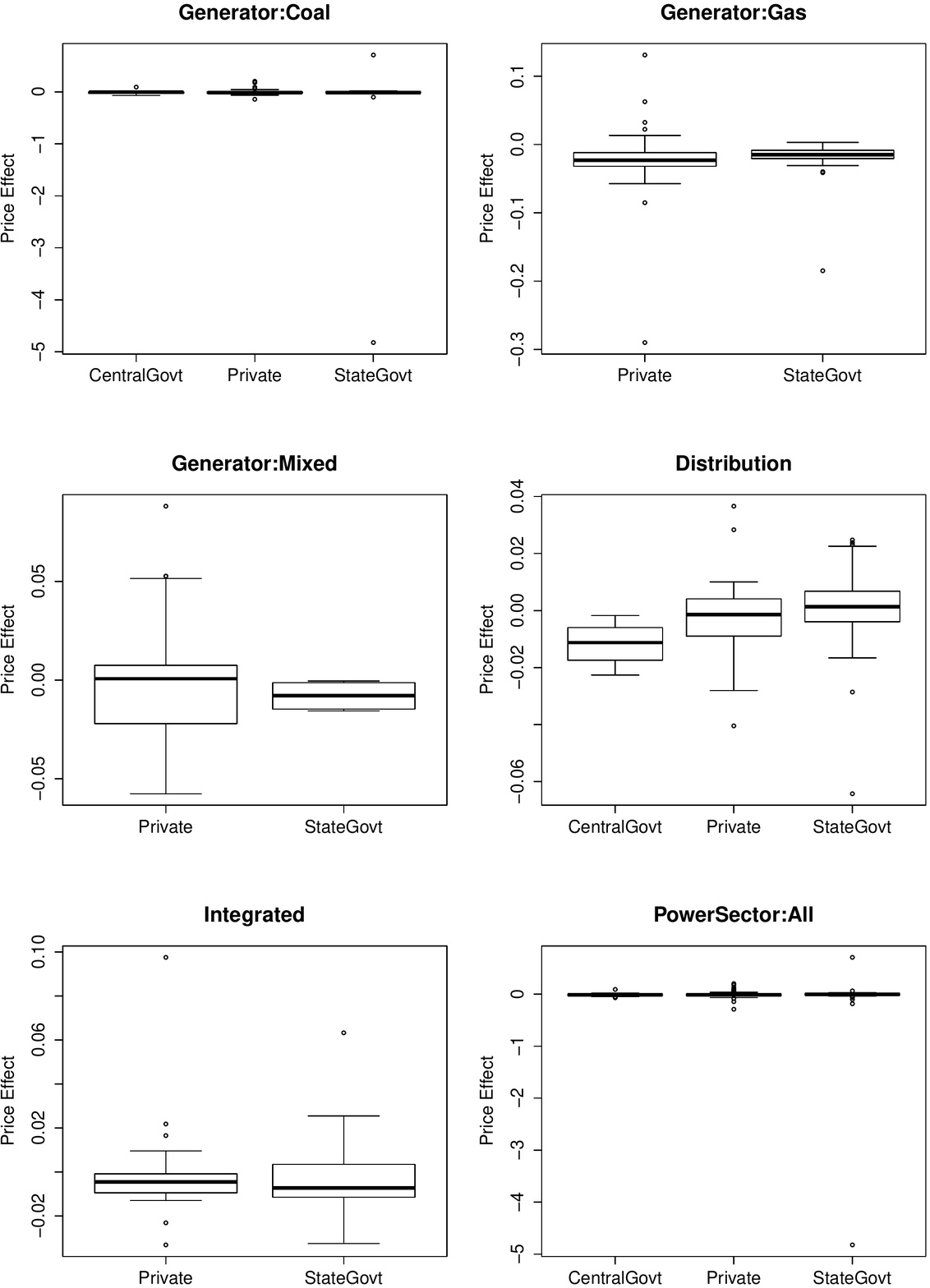}	
\end{figure}

\end{document}